\documentclass[
twocolumn,
superscriptaddress,
amsmath,amssymb,
aps,prb
]{revtex4-2}
\usepackage{appendix} 
\usepackage{graphicx}
\usepackage[caption=false,labelformat=simple]{subfig}
\usepackage{dcolumn}
\usepackage{braket}
\usepackage{threeparttable}
\usepackage{longtable}
\makeatletter
\def\LT@LR@e{\LTleft\z@   \LTright\z@}%
\makeatother

\usepackage{bm}
\usepackage{multirow}
\usepackage[ colorlinks, linkcolor=blue, anchorcolor=blue, citecolor=blue ]{hyperref}
\usepackage{cleveref}
\usepackage{verbatim}
\crefname{figure}{fig.}{figures}
\Crefname{figure}{Fig.}{Figures}
\crefname{subfigure}{fig.}{figures}
\Crefname{subfigure}{Fig.}{Figures}
\crefname{table}{table}{tables}
\Crefname{Table}{Table}{Tables}
\Crefname{equation}{Eq.}{Equations}

\begin{document}
%\title{Extended Hubbard corrected tight-binding model for rhombohedral multilayer graphene}
%\title{Self-consistent Extended Hubbard tight-binding model calculations for rhombohedral multilayer graphene}
%\title{Self-consistent Extended Hubbard calculations on tight-binding model Hamiltonians for rhombohedral multilayer graphene}
\title{Self-consistent tight-binding calculations with extended Hubbard interactions in rhombohedral multilayer graphene}
\author{Dongkyu Lee}
\affiliation{Department of Physics, University of Seoul, Seoul 02504, Korea}
\author{Wooil Yang}
\affiliation{Korea Institute for Advanced Study, Seoul 02455, Korea}
\author{Young-Woo Son}
\affiliation{Korea Institute for Advanced Study, Seoul 02455, Korea}
\author{Jeil Jung}
\email{jeiljung@uos.ac.kr}
\affiliation{Department of Physics, University of Seoul, Seoul 02504, Korea}

\date{\today}

\begin{abstract}
We study the mean-field broken symmetry phases of charge neutral multilayer rhombohedral graphene within tight-binding approximations including self-consistent extended Hubbard interactions.
We used on-site and inter-site Hubbard interactions obtained from a newly developed first-principles calculation method.
Our calculations for systems up to eight layers 
give rise to electron-hole asymmetries, band flatness, band gaps, and layer anti-ferromagnetic ground states in keeping with available experiments.
By including the intersite Hubbard interactions up to the next-nearest neighboring sites, the band gaps are shown to open when the number of layers is larger than three, while the trilayer system maintains its metallic nature with two low energy density of state peaks near the Fermi energy whose separation increases with the range of inter-site Hubbard parameters.
Within our framework, the calculated band gaps reflect mean-field ground states with extended Hubbard interactions, in closer agreement with experimental estimates. The tight-binding formulation further enables efficient treatment of large rhombohedral chiral systems, including twisted multilayer graphene.
\end{abstract}

\maketitle

\section{Introduction}\label{intro}
%
%1. Why RnG's are important.
%2. Why Extended Hubbard corrections in RnG's are important. 
Physical systems with nearly flat bands in momentum space have often spatially localized states in real space~\cite{rhim_singular_2021, park_quasi-localization_2024} that makes them susceptible to local Coulomb interactions. 
Example graphene systems with such nearly flat bands include the zigzag nanoribbons~\cite{wakabayashi_electronic_1999, son_energy_2006, jung_theory_2009, park_quasi-localization_2024},  
rhombohedral $n$-layer graphene (here we abbreviate as \textrm{RnG} and $n \ge 3$)~\cite{min_chiral_2008,pamuk_magnetic_2017, Kerelsky_moireless_2021, hagymasi_observation_2022},
and magic angle twisted bilayer graphene~\cite{choi_electronic_2019, koshino_maximally_2018, kerelsky_maximized_2019, bistritzer_moire_2011}
for which a variety of interaction driven phases have been studied in the literature.
%~\cite{pamuk_magnetic_2017, Kerelsky_moireless_2021, hagymasi_observation_2022}, the rhombohedral multilayer graphene on hexagonal boron nitride~\cite{chen_evidence_2019, chen_signatures_2019, gonzalez_topological_2021, park_topological_2023}, and the magic angle twisted bilayer graphene (MATBG)~\cite{choi_electronic_2019, koshino_maximally_2018, kerelsky_maximized_2019}. 
%
Recent advances in near-field optical identification of stacking domains of RnG have sparked a wave of research in RnG, in particular the tetralayer~\cite{liu_spontaneous_2023} and pentalayer~\cite{han_correlated_2024} systems where a variety of ordered phases have been observed, including the fractional quantum anomalous Hall effect, unconventional superconductivity.
We note that for these systems external system parameters such as the (proximity) spin-orbit coupling and gating are able to sensitively alter its properties~\cite{han_large_2024, lu_fractional_2024, zhou_fractional_2024, dong_theory_2024, choi_superconductivity_2025, waters_chern_2025, lu_extended_2025}.

From the perspective of modeling the RnG, the interplay between the band parameters and Coulomb interactions sensitively influences the onset of the ordered phases.
%and therefore it is %From the electronic structure calculation viewpoint it is desirable to capture properly the band Hamiltonian and the Coulomb interactions for a more reliable prediction of its properties.
%
% For bilayers
Whereas it is expected that an infinitesimally small Coulomb interaction strength is sufficient to trigger broken symmetry phases for $n \geq 2$ in a minimal $n$-chiral two-dimensional electron gas (2DEG) model~\cite{min_chiral_2008,zhang_spontaneous_2011},
actual calculation results sensitively depend both on the specific details in the band Hamiltonian and the Coulomb interaction model.
In particular, for Bernal stacked bilayer graphene, a number of studies predicted the possibility of various broken symmetry phases~\cite{lemonik_spontaneous_2010,vafek_many-body_2010,nandkishore_quantum_2010,zhang_spontaneous_2010,jung_lattice_2011}.
For instance, the Hartree-Fock calculations with long-range Coulomb interactions screened with a dielectric constant value  of $\varepsilon_r = 4$ gives rise to band gaps of the order of 20~meV for the bilayer system~\cite{jung_lattice_2011},
while the experimental signatures of Coulomb correlations~\cite{martin_local_2010, seiler_quantum_2022} 
and gaps in suspended bilayer graphene at charge neutrality turned out to be smaller than 2~meV~\cite{velasco_transport_2012, bao_evidence_2012}.
% 
%{\bf GJ: There are several other representative experimental measurements that should be cited and contextualized.}
% Trilayers 
The ABC stacked graphene or R3G is expected to have a larger band gap than the bilayer graphene within the same minimal model, although the actual gap magnitude turns out to be sensitive to details of the band's trigonal warping~\cite{jung_gapped_2013}.
In experiments, the R3G gaps have been reported to be between $0$ to $42$~meV~\cite{khodkov_direct_2015, van_elferen_fine_2013, bao_stacking-dependent_2011, lee_competition_2014, lee_gate-tunable_2022}. To achieve this gap within a $U$-only mean-field calculation we need $U = 4.8t$, where $t$ is the nearest-neighbor hopping parameter~\cite{lee_gate-tunable_2022}, which significantly surpasses the $U = 3.5t$ estimate of the constrained random phase approximation (cRPA)~\cite{wehling_strength_2011} and even the anti-ferromagnetic critical value of monolayer graphene $U = 2.2t$~\cite{sorella_semi-metal-insulator_1992}. 

On the computational front, gapless states were obtained for all RnG from first-principles calculations based on the density functional theory (DFT) with the local density approximation (LDA) or generalized gradient approximations (GGA) for exchange-correlation functional~\cite{adamo_toward_1999}, indicating the relative weakness of Coulomb interaction corrections in commonly used local and semi-local DFT approximations. 
On the contrary, inclusion of non-local Fock terms as in PBE0 hybrid functional allows to adequately estimate the 39~meV band gap for R3G~\cite{pamuk_magnetic_2017}, but overestimates the experimental gaps for ABCA stacked graphene or R4G of the order of $\sim 10$~meV~\cite{Kerelsky_moireless_2021, liu_spontaneous_2023}. 
Consequently, appropriate interaction models that predict results consistent with experiments for various RnG systems is needed. 
%It is desirable to propose for rhombohedral stacked n-layer graphene (RnG for $n=3, 4, \dotsb$) which has partially flat bands near the high-symmetry points $K$ and $K'$. 
%For example, results from density functional theory (DFT) local density approximation (LDA)~\cite{kohn_self-consistent_1965} and generalized gradient approximation (GGA)~\cite{perdew_generalized_1996} functionals fails to capture the band gaps in RnG systems. 
%that do not have strong enough short-local interactions, run the risk of predicting physics that deviate from their actual properties.
%
%computational point of view it is desirable to account refine the band Hamiltonian parameters to improve 
%Owing to the localized nature of the flat-band states, inclusion of electron-electron interactions lead to notable alterations in band flatness, band gap, correlation phase diagrams~\cite{Kerelsky_moireless_2021, guinea_electrostatic_2018, vahedi_magnetism_2021, gonzalez_magnetic_2021}. On the other hand, 

% --- Why good band parameters are desirable ---
%Recent experimental observations have underscored the need for more sophisticated calculations in flatband materials. The reported many-body phenomena in flatband, such as 

In this work, we obtain the interacting mean-field ground state of charge neutral RnG ($n=1, 2, \dotsb, 8$) based on the unified calculation framework of the self-consistent tight-binding approximations with the onsite ($U$) and intersite ($V$) Hubbard interactions (TB+$U$+$V$). 
Our main proposed model incorporates $U$ and $V$ from a newly developed {\it ab initio} DFT calculation method for the self-consistent extended Hubbard interactions (DFT+$U$+$V$)~\cite{lee_first-principles_2020,yang_ab_2021,Yang2024PRB}.
%and we obtain the hopping parameters from $\pi$-band maximally localized Wannier functions. 
We extracted nearest-neighboring and next nearest-neighboring intersite Hubbard interactions ($V_1$ and $V_2$) as well as onsite $U$ interactions from our {\it ab initio} calculations.
The calculated renormalized hopping parameters within TB+$U$+$V$ provide a Fermi velocity ($v_F$) of single layer graphene that is enhanced by almost 30\% with respect to the computed value using DFT-LDA~\cite{jung_tight-binding_2013} for our choice of cutoff range up to $V_2$. We note that the $v_F$ sensitively depends on the dielectric screening from  substrates, e.g., it is larger than the DFT-LDA velocity by 10$\sim$30\% if deposited on top of SiO$_2$~\cite{dorgan_mobility_2010, knox_spectromicroscopy_2008,hwang_fermi_2012} and up to almost 50\% when deposited on top of hBN~\cite{yu_interaction_2013, muzzio_momentum-resolved_2020, zhang_experimental_2005, hwang_fermi_2012}.
%The ground states of the TB+$U$+$V$ Hamiltonian are able to capture the particle-hole asymmetry, antiferromagnetic band gap, flat bandwidth, and ordering critical temperatures for $n \geq 4$. 
% --- --- ---
Using a newly developed computational method, we aim to provide a more realistic description of the band structure that incorporates the band renormalization due to Coulomb interaction effects in RnG systems. Within our TB+$U$+$V$ framework, our calculated band gaps correspond to the truncated-range mean-field Hartree-Fock ground states, that are more closely comparable to available experimental estimates.
%measurements as we will show later on. 
Additionally, the TB-based description of the bands allows for a more flexible and efficient calculation framework that can eventually be used to study larger number of atoms with rhombohedral stacking orders such as twisted multilayer graphene systems~\cite{gonzalez_magnetic_2021,Nakatsuji2023PRX,Park2025Nature}.  
%
%Calculations on R3G for several Coulomb interaction ranges provide alternative viewpoints for the various reported experimental gaps depending on the screening environment due to substrate choice. 
%and charge density wave~\cite{chen_signatures_2019, cao_unconventional_2018, bhowmik_broken-symmetry_2022, tseng_anomalous_2022, han_large_2023}, 
%{\bf GJ: Need to motivate better the significance of the findings reported in this manuscript}

\section{Self-consistent tight-binding extended Hubbard model}\label{sec2}

In this section we first briefly introduce a DFT+$U$+$V$ method to obtain self-consistent extended Hubbard interactions. Then, we present our newly developed TB+$U$+$V$ method for RnG systems where efficient calculations are possible thanks to the reduced Hilbert space of the tight-binding orbital basis.

\subsection{DFT+$U$+$V$ for $U$, $V$ parameters calculations}\label{sec2-0}
In typical DFT approaches with Hubbard interactions, the $U$ and $V$ parameters are found by empirical fitting procedures~\cite{Anisimov1991PRB, Anisimov1997JPC}. Recently, however, there have been significant developments in computing those parameters {\it ab initio}~\cite{Kulik2006PRL, Cococcioni2005PRB, Miyake2008PRB, Miyake2009PRB, Aichhorn2009PRB, Mosey2007PRB, Mosey2008JCP, agapito_reformulation_2015, Rubio2017PRB, Timrov2018PRB, lee_first-principles_2020, tancogne-dejean_parameter-free_2020, Timrov2021PRB,yang_ab_2021,Yang2024PRB}. 
Among them, we used a first-principles calculations based method~\cite{lee_first-principles_2020,yang_ab_2021,Yang2024PRB} 
developed by extending to $V$ parameters the Agapito–Curtaolo–Buongiorno Nardelli pseudohybrid functional for on-site Coulomb $U$ interactions~\cite{agapito_reformulation_2015}. 
This method turns out to be very efficient and accurate for obtaining various physical parameters such as bands gaps, atomic forces, phonon dispersions and magnetic moments of correlated solids~\cite{lee_first-principles_2020,yang_ab_2021,Yang2024PRB,tancogne-dejean_parameter-free_2020, Jang2023PRL}. 
Moreover, this method can self-consistently determine the strength of inter-site Hubbard interactions between a pair of orbitals with arbitrary spatial range to handle %long-ranged
local and nonlocal Coulomb interactions in low dimensional solids~\cite{lee_first-principles_2020}. 
An advantage of our implementation with respect to other mean-field Hubbard model calculations~\cite{hubbard_electron_1963, fernandez-rossier_magnetism_2007} is that the extended Hubbard $V$ parameters are determined self-consistently.  We refer the previous related studies~\cite{lee_first-principles_2020,yang_ab_2021,Yang2024PRB,tancogne-dejean_parameter-free_2020, Jang2023PRL} for further details on self-consistent calculations of extended Hubbard interactions. 

\subsection{Self-consistent TB+$U$+$V$ calculations}\label{sec2-1}

\begin{figure}[b] %>>>> FIG 1 >>>>% 
\includegraphics[width=0.90\columnwidth]{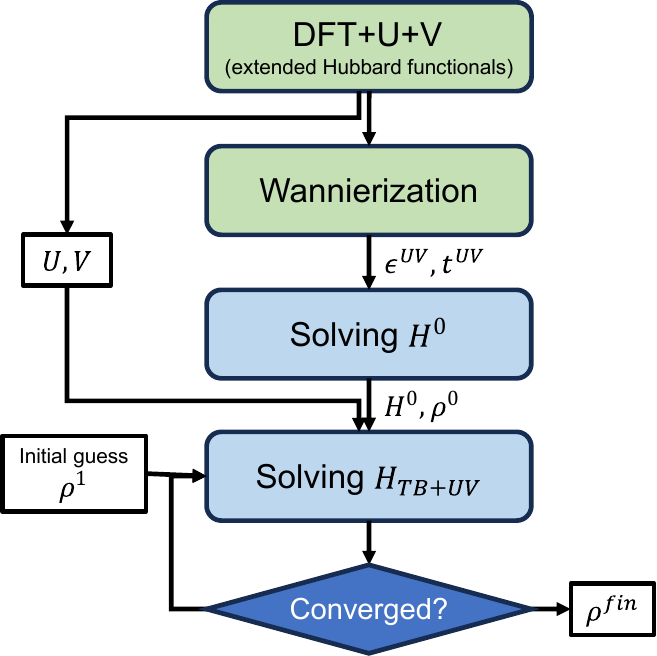}
\caption{A flowchart diagram summarizing the construction and solution process of the TB+$U$+$V$ model. The hopping parameters of the TB band Hamiltonian are extracted from the DFT+$U$+$V$, where the $U$ and $V$ parameters are obtained in a self-consistent manner~\cite{lee_first-principles_2020,yang_ab_2021,Yang2024PRB},
and whose associated ground state density is $\rho^0$. 
The degrees of freedom associated with $U$ and $V$ are eliminated using the extended Hubbard functional method.}\label{Fig:fig1} 
\end{figure} %<<<< FIG 1 END <<<<%

%The DFT$+UV$ calculations%~\cite{lee_first-principles_2020} give rise to the Kohn-Sham (KS) ground states and the extended Hubbard on-site $U$ and inter-site $V$ term values.
In the following, we explain the calculation procedure of our self-consistent TB+$U$+$V$ bands
in search of broken-symmetry solutions starting from the reference Hamiltonian resulting from the
DFT+$U$+$V$ calculations. 
%{\bf We will neglect the role of $V_{\perp}$, while we verified it is worthwhile to maintain up to the second nearest neighbor in-plane $V_2$.} 
%and that a step to solve tight-binding Hamiltonian of $H^0$ is added subsequently. 
%We note that this approach also helps prevent the double-counting of interactions and enhances accuracy significantly. 
\Cref{Fig:fig1} shows a diagram explaining the construction process and resolution of $\hat{H}^{{\rm TB}+U+V}$ Hamiltonian.
The procedure is as follows. We start with the DFT+$U$+$V$ calculations to obtain the corresponding tight-binding Hamiltonian $\hat{H}^0$, its ground state density matrix $\rho^0$, the associated tight-binding and the extended Hubbard Coulomb interaction parameters. 
Through the DFT+$U$+$V$ calculations we obtain self-consistently the tight-binding hopping terms and the Hubbard $U$ and $V$ parameters. 
We have carried out valley, spin, and inversion symmetry preserving self-consistent DFT+$U$+$V$ calculations to provide the reference renormalized tight-binding Hamiltonian $\hat{H}^0$.
Then, this symmetry preserving ground-state of $\hat{H}^0$ is taken as the reference starting point of the self-consistent TB+$U$+$V$ calculations. At this stage we seek for broken symmetry solutions  where we use an initial condition a density matrix $\rho^{1}$ with broken symmetry until we converge the calculations to self-consistency.

The Kohn-Sham (KS) equations of a DFT+$U$+$V$ calculation can be cast onto a tight-binding Wannier orbital basis that provides the reference point Hamiltonian $\hat{H}_0$ of our self-consistent TB+$U$+$V$ calculation,
\begin{align} \label{Eq:eq_H0}
\hat{H}^{0} &= \sum_{i\sigma} h^{\rm KS}_{ij} \hat{c}^\dagger_{i\sigma} \hat{c}_{j\sigma} = \sum_{i j \sigma} 
\epsilon^{UV}_{i} \hat{c}^\dagger_{i\sigma} \hat{c}_{i\sigma}  +\sum_{ij\sigma} t^{UV}_{ij} \hat{c}^\dagger_{i\sigma} \hat{c}_{j\sigma},
\end{align}
where $\epsilon^{UV}_i$ is the onsite potential of the $i$-th site and $t^{UV}_{ij}$ the hopping parameters
between $i$- and $j$-th sites.
For simplicity, all our calculations have started from the non-magnetic spinless form of $\hat{H}^0$
and use abbreviated notations for the density matrix $\rho^{\sigma \sigma^{\prime}}_{ij} = \langle \hat{c}^{\dag}_{i \sigma} \hat{c}_{j \sigma^{\prime}} \rangle$ of the form $\rho_{i\sigma} = \rho^{\sigma \sigma}_{ii}$ and $\rho_{i}=\rho_{i\uparrow}+\rho_{i\downarrow}$ by removing repeated site indices and spins.
The $\epsilon^{UV}$ and $t^{UV}$ are renormalized by the extended Hubbard Coulomb interactions, that removes the double counting of the Coulomb interactions implied in DFT+$U$+$V$~\cite{Cococcioni2005PRB}:
\begin{align} 
\epsilon^{UV}_{i\sigma}  &= \epsilon^{\text{NI}}_{i\sigma} + \sum_{j\neq i} V^H_{ij} \rho^0_{j}  + v_{xc, \,i \sigma} + U_i \rho^0_{i\tilde{\sigma}}~, \label{eq:onsite_UV} \\ 
t^{UV}_{ij\sigma}  &= t^{\text{NI}}_{ij\sigma} - V_{ij} \rho^0_{ij\sigma} \label{eq:t_UV},
\end{align}
%where the terms of the KS Hamiltonian and 
where the $\rho^{0}_{ij}$ density matrix elements corresponds to the DFT+$U$+$V$ ground-state in \Cref{Eq:eq_H0}, $\epsilon^{\text{NI}}$ and $t^{\text{NI}}$ represent the non-interacting parts of the tight-binding parameters. See \Cref{Appendix_TBUV} for further details. 
The second term on the right hand side of \Cref{eq:onsite_UV} is the Hartree term proportional to the bare Coulomb potential $V^H_{ij}$ between sites $i$ and $j$. The third is the exchange-correlation potential at site $i$ for spin $\sigma$. The last terms of \Cref{eq:onsite_UV} and \Cref{eq:t_UV} are the extended Hubbard corrections where $U_i$ and the $V_{ij}$ parameters are obtained from the DFT+$U$+$V$ calculation.

% --- Definition of reference starting Hamiltonian ---
%Our self-consistent calculations 
%Hamiltonian can be defined for small perturbations such as different $k$-mesh, temperature, doping, and gating. 
Our self-consistent TB+$U$+$V$ calculations take as starting point reference the interaction-less tight-binding Hamiltonian,
\begin{align} 
\hat{H}^{{\rm TB}0}[\rho_0] =& \hat{H}^{0} - \hat{H}^{UV}[\rho_0],
\end{align}
where we remove the $U+V$ terms included in $\hat{H}^0$.
%in Eq.~(\ref{Eq:eq_H0}).
% 
%in search of lower energy broken-symmetry states.
%Successive iterations are carried out to find the converged density matrix $\rho$ that satisfies the equation  
%The interaction correction involves excluding the original electron-electron interaction from the Hamiltonian and introducing a new one based on the updated density matrix. In other words, the interaction corrected Hamiltonian used in the $(N+1)$th step,  density matrix obtained from the $N^{th}$ self-consistent iteration step is $\rho^N$, is given by,
%\begin{equation}
%%H_{{\rm TB}+U+V}\left[ \rho \right] = H^{0} + H^{UV}\left[ \rho-\rho^0 \right], \label{selfcon}
%\hat{H}^{{\rm TB}+U+V}\left[ \rho \right] = \hat{H}^{{\rm TB}0}[\rho_0] %- \hat{H}^{UV}\left[\rho^0 \right] 
%+ \hat{H}^{UV}\left[\rho \right], \label{selfcon}
%\end{equation} 
%
The self-consistent $\rho$ dependent TB+$U$+$V$ Hamiltonian is
\begin{align} 
\hat{H}^{{\rm TB}+U+V}\left[ \rho \right] %= & \hat{H}^0 - \hat{H}^{UV}[\rho_0] + \hat{H}^{UV}\left[\rho \right] \\
=& \hat{H}^{{\rm TB}0}[\rho_0] 
+ \hat{H}^{UV}\left[\rho \right] ,  \label{selfcon}
%= & \hat{H}^0  +  \hat{H}^{UV}\left[ \rho - \rho_0\right]  
\end{align} 
where the $U+V$ correction is given by 
\begin{align*} 
\hat{H}^{UV}\left[ \rho \right] =& \sum_{i\sigma} (U_i \rho_{i\tilde\sigma} + \sum_{j\neq i} V^H_{ij} \rho_{j}) \hat{c}^\dagger_{i\sigma} \hat{c}_{i\sigma} \\
&- \sum_{ij\sigma}V_{ij} \rho_{ij\sigma} \hat{c}^\dagger_{i\sigma} \hat{c}_{j\sigma}.~ 
\end{align*} 
Here in $\hat{H}^{UV}[\rho]$ we have neglected the corrections proportional to $v_{xc}$ in ~\Cref{eq:onsite_UV}, and the equation would reduce to the truncated-range Hartree-Fock form if $V^{H}_{ij} = V_{ij}$,
but in general they can be defined differently such that $V^{H}_{ij} \neq V_{ij}$. 
We note that the Hamiltonians in \Cref{Eq:eq_H0} and \Cref{selfcon} are the same if $\rho = \rho^0$,
namely when the self-consistent TB+$U$+$V$ calculation does not develop a new solution.

\subsection{Rhombohedrally stacked n-layer graphene}\label{rhomboG}

\begin{table}[bt!] %>>>> TABLE >>>>%
\begin{threeparttable}
\caption{
Calculated in-plane extended Hubbard parameters $U$, $V_1$ and $V_2$ and the Fermi velocity of single layer graphene~(R1G) in the DFT+$U$+$V$ calculation and comparison with other methods.
The Hubbard parameters are given in $eV$ units. 
}\label{tab:table1}
\begin{ruledtabular}
\def\arraystretch{1.2}%  1 is the default, change whatever you need
\begin{tabular}{ccccc}
 & This work & ref\tnote{a} & ref\tnote{b} & ref\tnote{c}\\
\hline
$U$ & 6.20 & 7.56 & 10.16 &-\\
$V_1$ & 3.22 & 4.02 & 5.68 &-\\
$V_2$ & 2.09 & 2.57 & 4.06&-\\
\hline
$v_{\text{f}}$~[$10^6 m/s$] & 1.10 & 1.43 & -& 0.84 \\
$t_{\text{eff}}$~[$eV$] & $-$3.42 & $-$4.46 & -& $-$2.58 \\
\end{tabular}     
\begin{tablenotes}
\item [a] Extended Hubbard functional DFT+$U$+$V$~\cite{tancogne-dejean_parameter-free_2020}
\item [b] cRPA~\cite{schuler_optimal_2013}
\item [c] LDA~\cite{jung_tight-binding_2013}
\end{tablenotes}
\end{ruledtabular}
\end{threeparttable}
\end{table} %>>>> TABLE-END >>>>%

\begin{figure}[bth] %>>>> FIG 2 >>>>%
\subfloat[\label{Fig:fig2_a}]{\includegraphics[width=0.90\columnwidth]{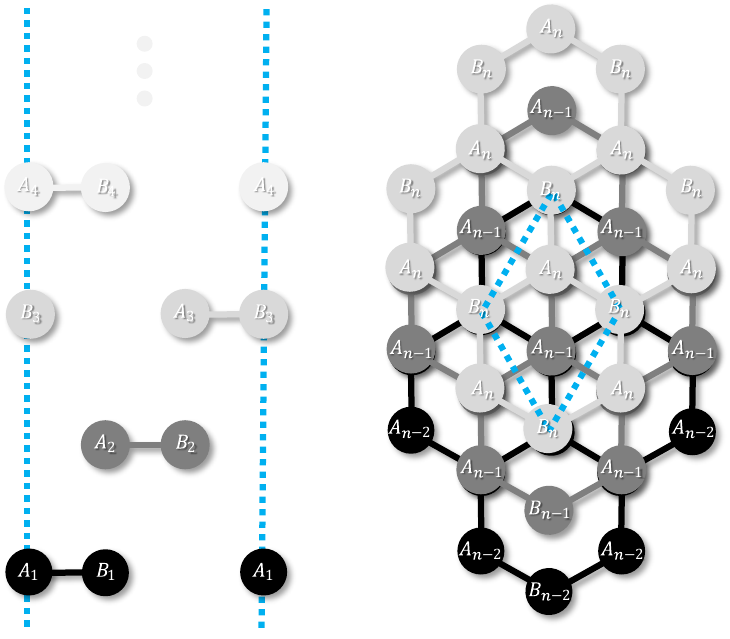}}\\
\subfloat[\label{Fig:fig2_b}] {\includegraphics[width=0.90\columnwidth]{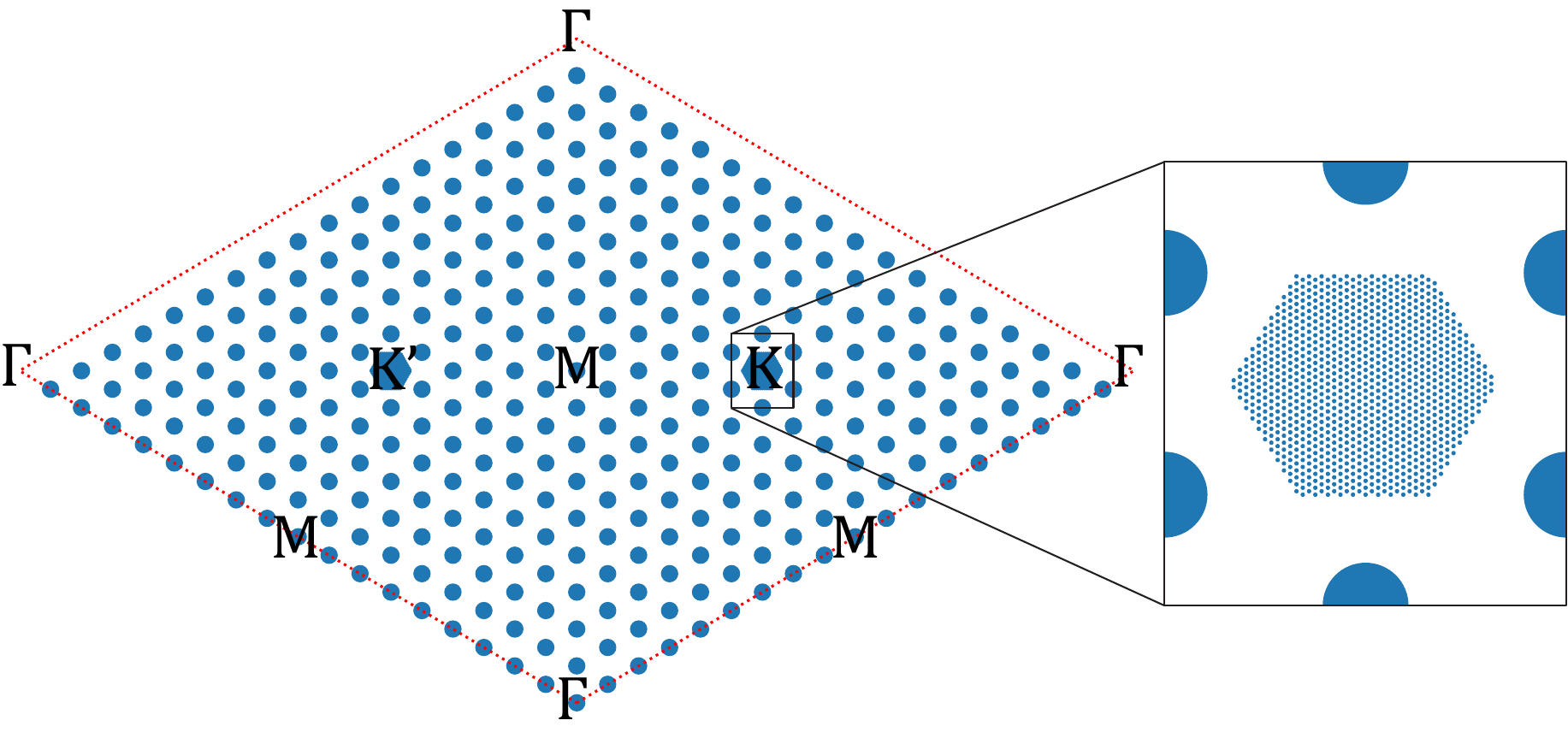}} 
\caption{(a)~Schematic sideview(left) and topview(right) of rhombohedral stacked few-layer graphene structure. The label $A_i$ or $B_i$ ($i = 1, 2, \dotsb, n$) represents the $A$ or $B$ sublattice orbital of the $i$-th layer. The dashed lines denote the unit cell of the RnG. (b)~An example of the partially dense $k$-point sampling used to solve our TB+$U$+$V$ Hamiltonians of the RnG. The points at $K$($K'$) position are replaced by dense grids. The size of each point corresponds to its weight. The illustrated example consists of a $18 \times 18$ coarse grid with $32 \times 32$ partially dense points resulting in an effective sampling of 576 × 576 points near the valleys.}\label{Fig:fig2}
\end{figure} 

We carried out the TB+$U$+$V$ calculations for RnG which are systems with partially flat bands. We have considered in this work rigid structures of which the in-plane lattice constant is $a=2.46 \text{~\AA}$ and the interlayer distance is $d=3.35 \text{~\AA}$. \Cref{Fig:fig2_a} illustrates the unit cell of rhombohedral multilayer graphene. The DFT calculations were performed with modified {\textsc{Quantum ESPRESSO}}~\cite{giannozzi_quantum_2009,lee_first-principles_2020,yang_ab_2021} for the extended Hubbard functional DFT+$U$+$V$. The codes can be found on github~\cite{dftuv_github}. 
We used a projector augmented wave (PAW)~\cite{blochl_projector_1994} LDA pseudopotential parameterized by Perdew and Zunger~\cite{perdew_self-interaction_1981} in {\textsc{PSlibrary}}~\cite{dal_corso_pseudopotentials_2014}. The Brillouin zone integrations in this step were performed with $60 \times 60 \times 1$ Monkhorst-Pack mesh points. In these calculations, we have introduced a extended Hubbard correction cutoff of $L_{UV} = 2.46~\text{\AA}$ that gives finite $V_1$ and $V_2$ values.
The onsite parameter $U$, nearest inter-sublattice parameter $V_1$, and nearest intra-sublattice parameter $V_2$ were self-consistently determined for this interaction range. 
The calculated Hubbard parameters were similar for the different systems considered and thus have used the same values in all cases.  
%we did not find significant differences.
%depending on the layer, sublattice, or structure, 
%the same values were used in all cases.
%for all bases and for every RnG. 
In our calculations we did not consider the inter-layer Coulomb interaction $V$ terms.
The DFT+$U$+$V$ tight-binding parameters $\epsilon^{UV}$ and $t^{UV}$ were obtained by constructing the maximally localized Wannier functions (MLWF) using {\textsc{Wannier90}}~\cite{pizzi_wannier90_2020}. The obtained in-plane hopping parameters were truncated with the $F_2G_2$ method~\cite{jung_accurate_2014} up to the second same-sublattice nearest neighbors. 
Likewise, we keep the interlayer hopping parameters up to the second nearest out-of-plane layers.
%between The hopping terms between the basis differing by three or more layers were excluded, as they were negligible in magnitude. 
The truncation method and the truncated hopping models for RnG band Hamiltonian are explained in~\Cref{Appendix_FnGn}.

\Cref{tab:table1} summarizes the parameters obtained during the DFT+$U$+$V$ step for single layer graphene. Our Hubbard parameters are smaller than similar earlier calculations~\cite{tancogne-dejean_parameter-free_2020}
or the cRPA~\cite{schuler_optimal_2013}, likely due to differences in implementation details such as $k$-point grid density, resulting also in a smaller predicted Fermi velocity.
%Due to $\pi$-bonding, the $\pi$-band orbitals exhibit a broader spread compared to the $p_z$ orbital, and this spread is calculated to increase with the density of the reciprocal space grid~\cite{jung_tight-binding_2013}. 
%As the spread of orbitals increases, it reduces the magnitude of local interactions. 
%Our results yield smaller values compared not only to cRPA but also to the same extended Hubbard functional DFT+$U$+$V$ method. 
The Fermi velocity $v_\textrm{F}$ obtained using the slope of the Dirac cone in monolayer graphene (R1G) is larger than the LDA velocity~\cite{jung_tight-binding_2013, jung_accurate_2014}, and in reasonable agreement with $\upsilon_{\rm F} \sim 1.05 \cdot 10^6$m/s commonly used in modeling of graphene on SiO$_2$~\cite{dorgan_mobility_2010, knox_spectromicroscopy_2008,hwang_fermi_2012} and somewhat smaller than $\upsilon_{\rm F} \sim 1.2 \cdot 10^6$m/s used in graphene on hBN~\cite{yu_interaction_2013, muzzio_momentum-resolved_2020, zhang_experimental_2005, hwang_fermi_2012}. The effective hopping value $t_{\textrm{eff}}$ calculated numerically from the Fermi velocity is also enhanced over the LDA calculation. We also observed that as $L_{UV}$ increases, the Fermi velocity becomes larger (e.g., $v_F = 1.25 \cdot 10^6$~m/s for $L_{UV} = 3a$). This behavior is related to the logarithmic divergence of the Fermi velocity in graphene, which arises due to long-range exchange interactions~\cite{elias_dirac_2011}.

The final step in the construction of the TB+$U$+$V$ Hamiltonian for RnG involved solving the $H^0$, which includes the hopping parameters, to obtain $\rho^0$. (The energy dispersions near the $K$ point for the non-corrected Hamiltonians can be found in \Cref{fig:figB2}.) 
For the Hartree term, we adopt the bare Hartree potential~\cite{min_ab_2007, jung_enhancement_2011} 
\begin{equation}\label{eq:Hartree}
  V^H_{ij} = \frac{1}{N_{ij}A}\sum_{\bm G}e^{i \bm G \cdot(\tau_i-\tau_j)}|f(|\bm G|)|^2V_q(|\bm G|)  
\end{equation}
where $N_{ij}$ is the number of nearest $j$ orbitals with respect to $i$ orbital, $A$ is the system area, and $\tau_i$ is the position of $i$ orbital in the unitcell. In our calculations, we have used the bare Coulomb interactions $V_q(q) = 2\pi e^2/q$ for intra-layer interaction, $V_q(q) = 2\pi e^2\textrm{exp}(-qd)/q$ for inter-layer interaction and the form factor $f(q) = [1-(r_0q)^2]/[1+(r_0q)^2]^4$ with $r_0 = a/(6\sqrt{10})~\text{\AA}$, a somewhat extended radial distribution of the 2p electrons following ~\cite{jung_enhancement_2011}. 
%
%{\bf GJ: Tobias Stauber proposed form factor with more realistic radius, let us discuss. This term is of importance for capturing the electronic screening in the presence of potentials that produce charge imbalance like perpendicular electric fields or in systems with ionic bonds.} 
%
For RnG at charge neutrality, the details in modeling the interlayer Hartree potential is not critical, while it is expected to be more important in systems with a perpendicular electric field.
%{\bf GJ Let us talk about this too. As long as the Hartree potential is not excessively large and remains within a reasonable range, the convergent states were identity. In contrast, in our tests on biased bilayer graphene, we found that the $\bm G = 0$ term must be set to twice the previously reported value~\cite{min_ab_2007} in order to reproduce the same results with DFT(+$U$+$V$). For a detailed explanation regarding the Hartree term in the biased systems, please refer to \Cref{Appendix_Hartree}.}

During the self-consistent TB+$U$+$V$ Hamiltonian calculation, adaptive $k$-point sampling was employed to resolve details near the Dirac points, as illustrated in~\Cref{Fig:fig2_b}. This approach involves creating a coarse grid of dimensions $N_\textrm{coarse}\times N_\textrm{coarse}\times 1$ across the entire Brillouin zone (BZ) and replacing the representative zone at the $K$ point with a dense grid of dimensions $N_\textrm{dense}\times N_\textrm{dense}\times 1$. In this paper, we used $N_\textrm{coarse}=24$ and $N_\textrm{dense}=64$, resulting in an effective $k$-grid density of $1536 \times 1536 \times 1$ near the Dirac points. The density matrix was mixed using the modified Broyden algorithm~\cite{johnson_modified_1988}, and iterations were continued until the distance between the normalized  eigenstate vectors between successive steps was below $10^{-6}$.

\section{Low energy nearly flat bands and Coulomb-interaction driven gaps}\label{sec3}

\begin{figure*}[tbh!] %>>>> FIG 3 >>>>% 
\subfloat[\label{Fig:fig3_a}]{\begin{tabular}{c}
\includegraphics[width=0.325\textwidth]{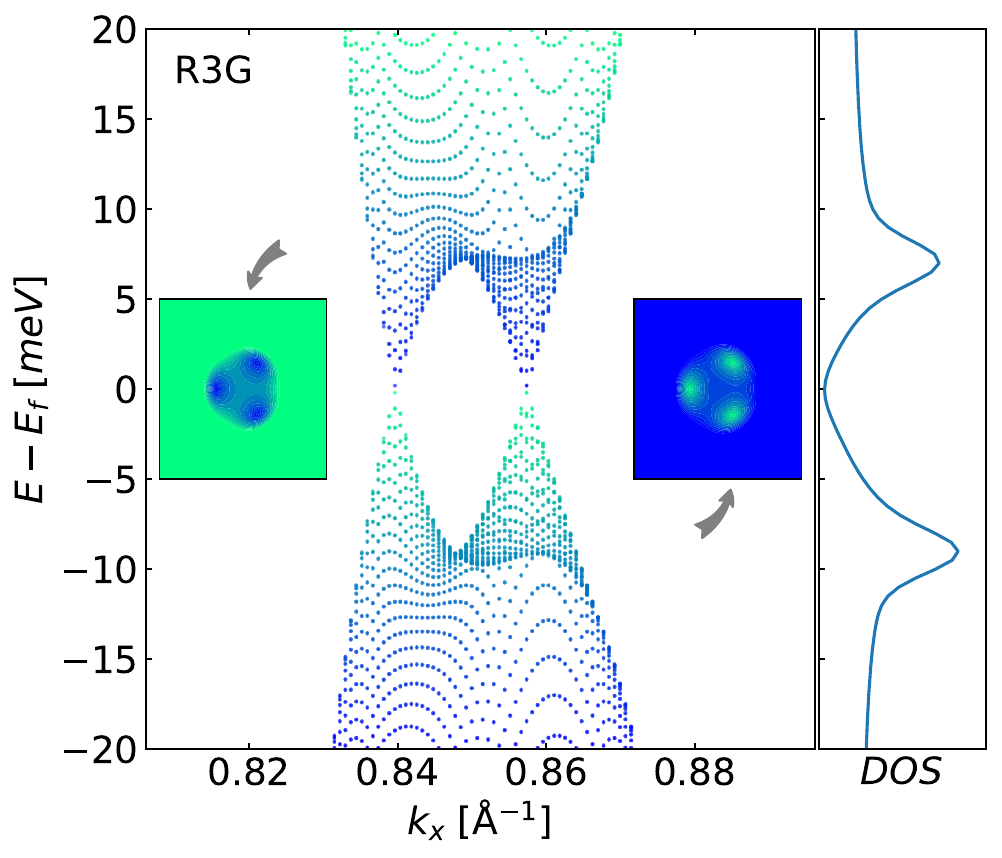}
\includegraphics[width=0.325\textwidth]{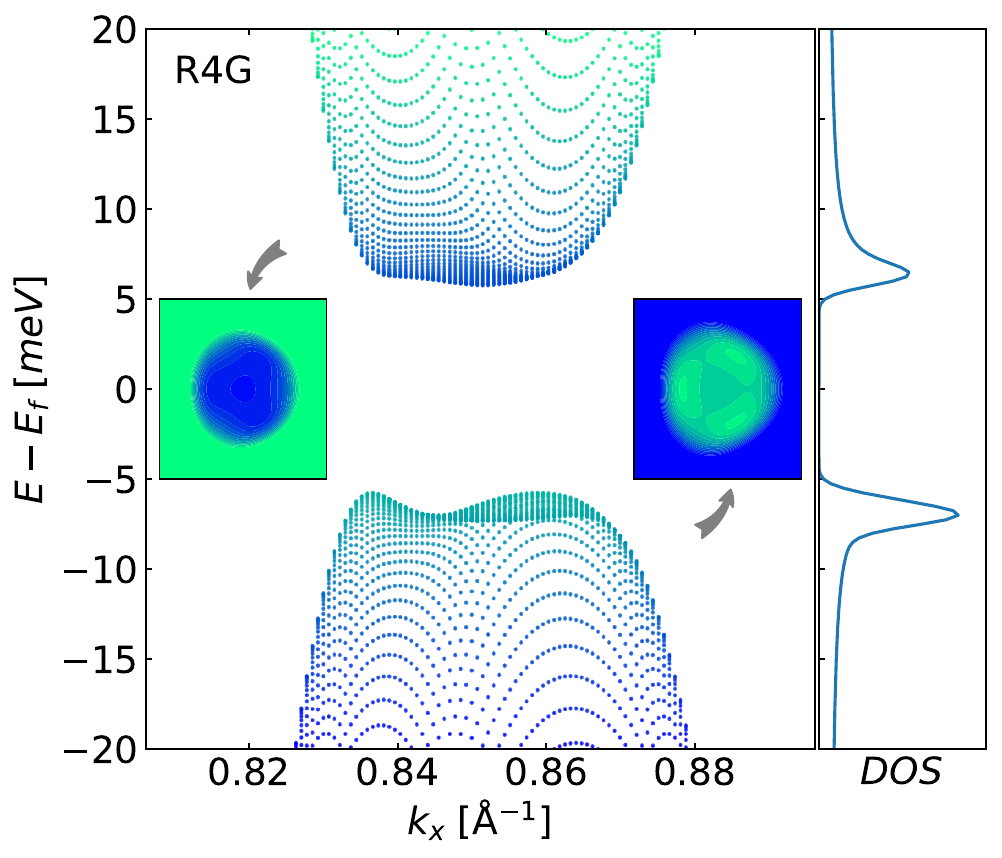}
\includegraphics[width=0.325\textwidth]{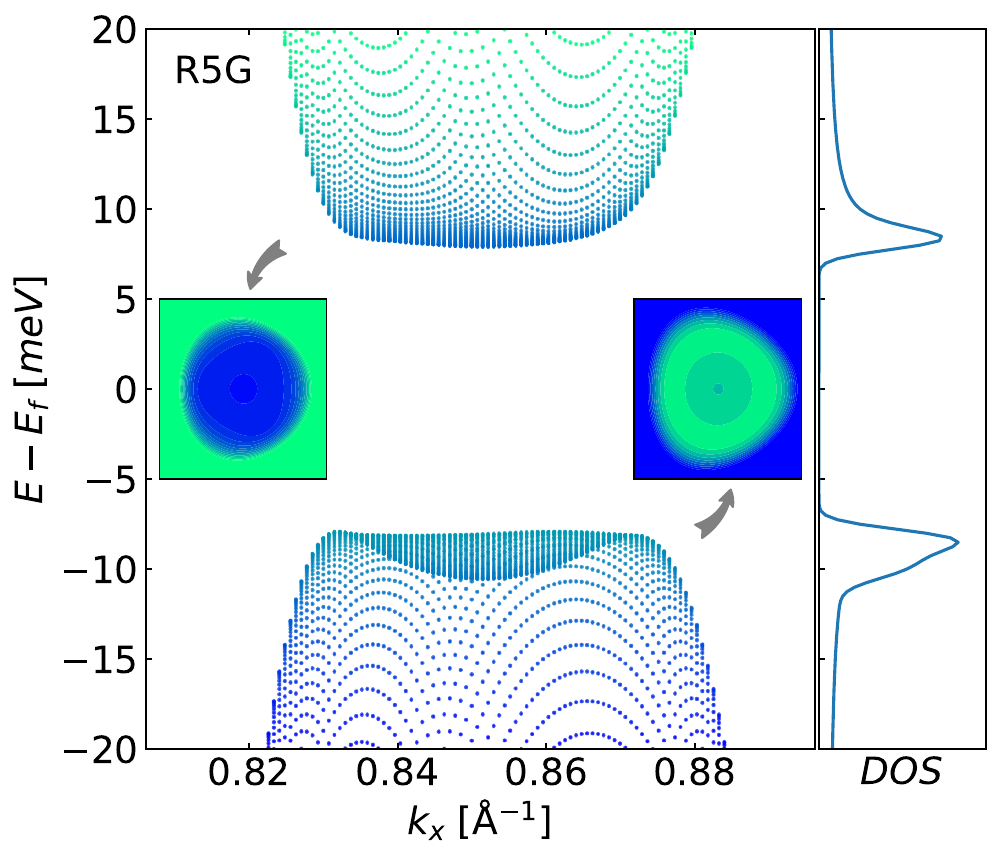}\\
\includegraphics[width=0.325\textwidth]{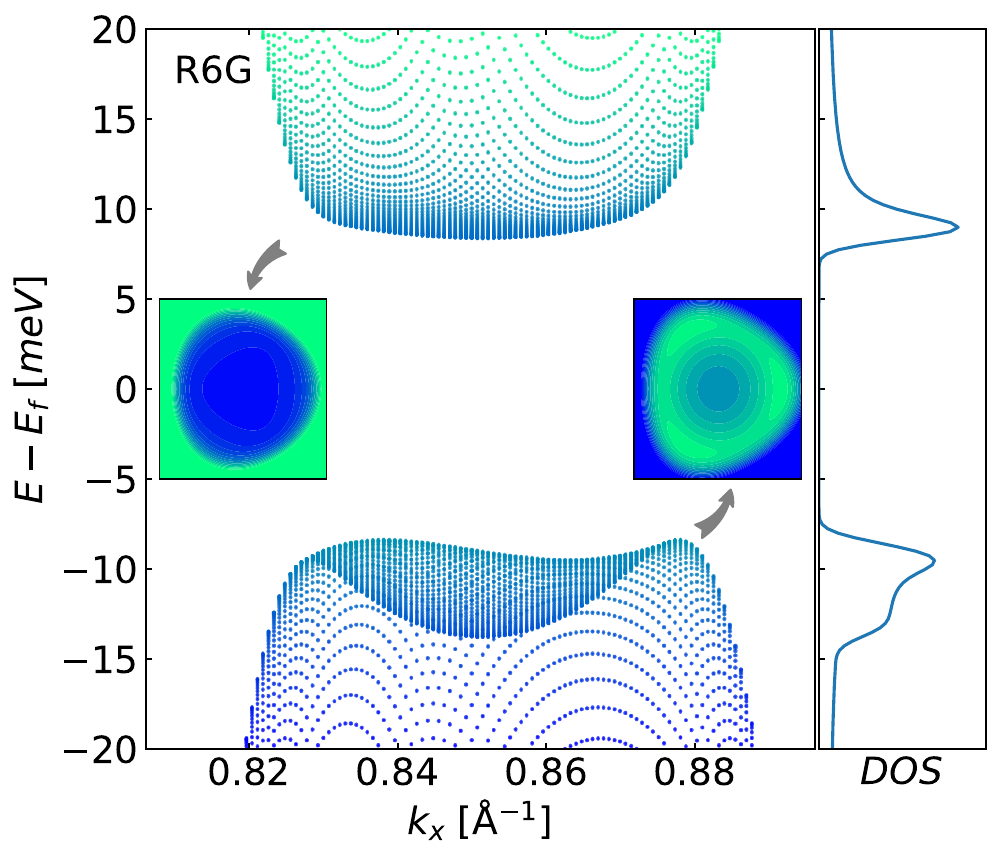}
\includegraphics[width=0.325\textwidth]{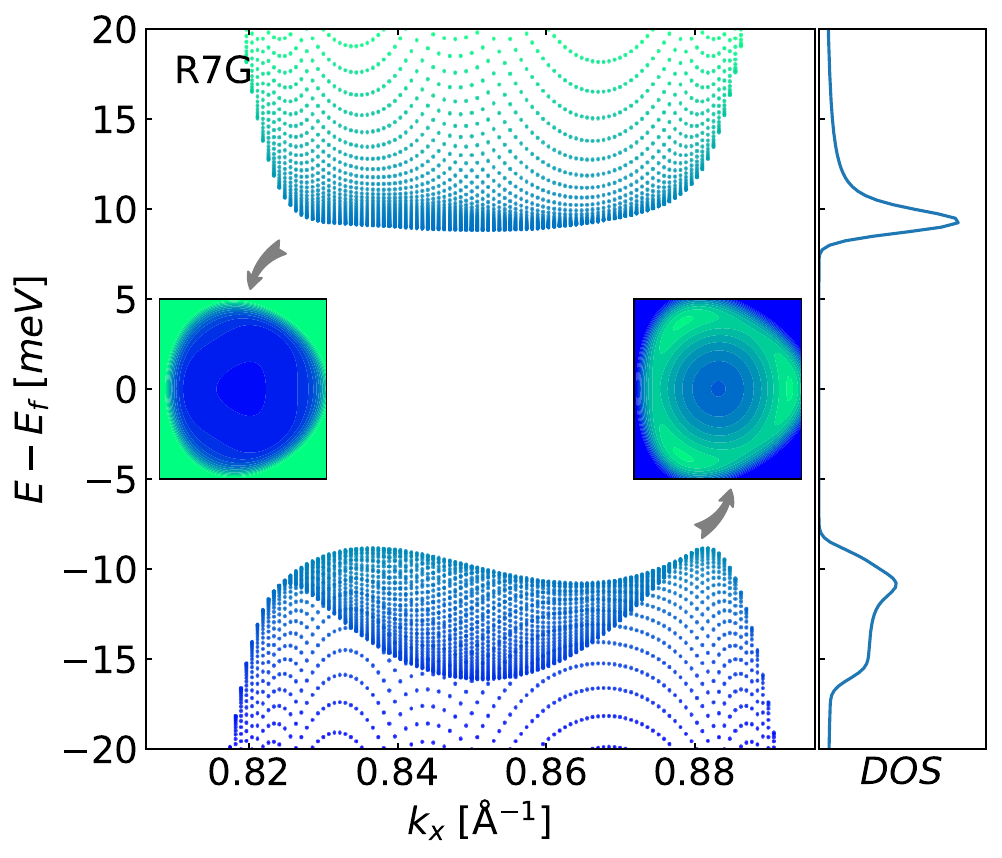}
\includegraphics[width=0.325\textwidth]{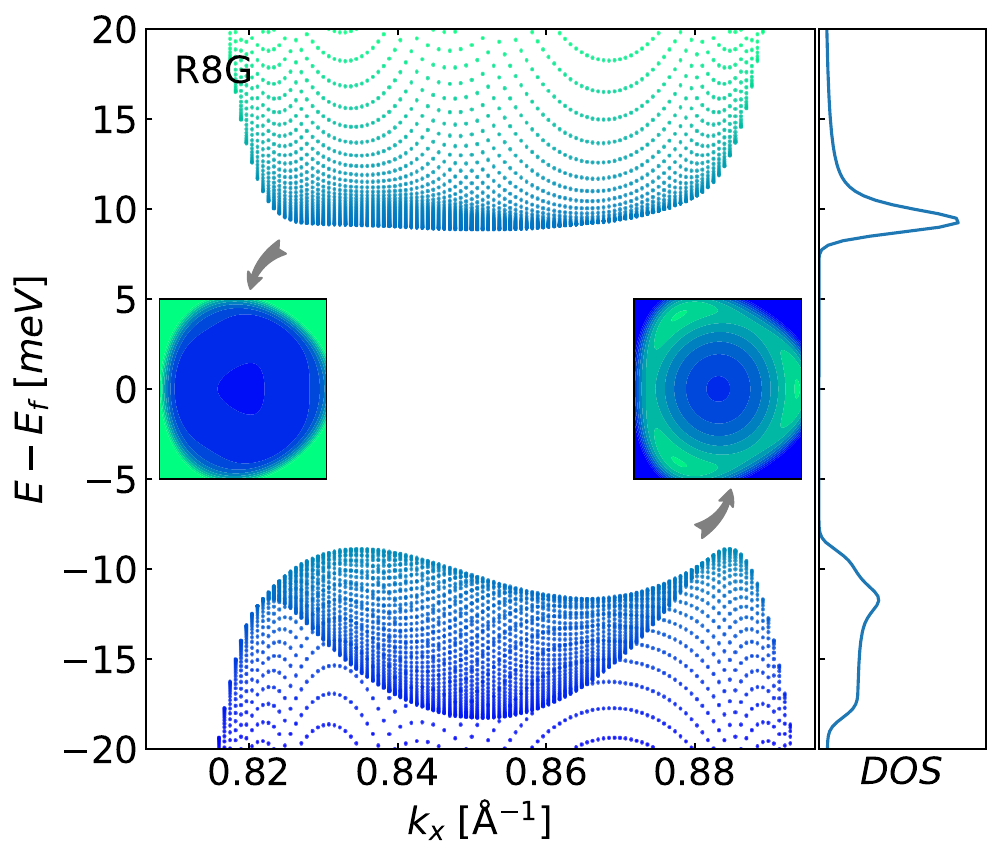}
\end{tabular}}\\
\subfloat[\label{Fig:fig3_b}]{\includegraphics[width=0.325\textwidth]{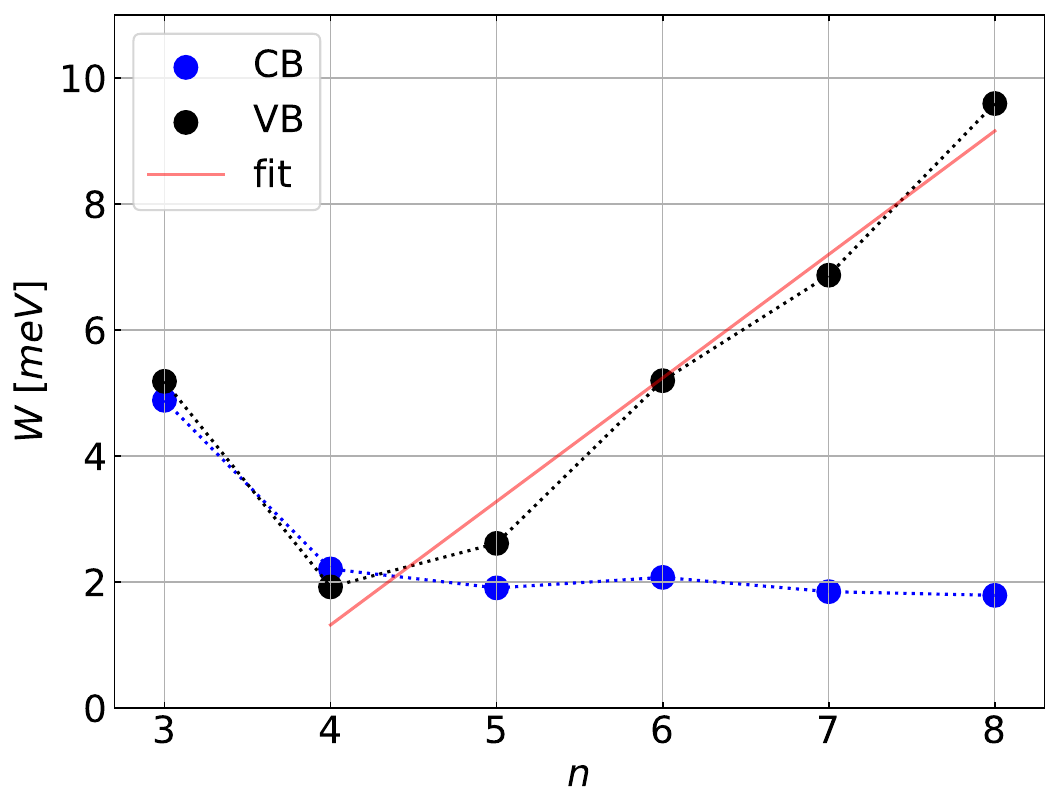}}
\subfloat[\label{Fig:fig3_c}] {\includegraphics[width=0.325\textwidth]{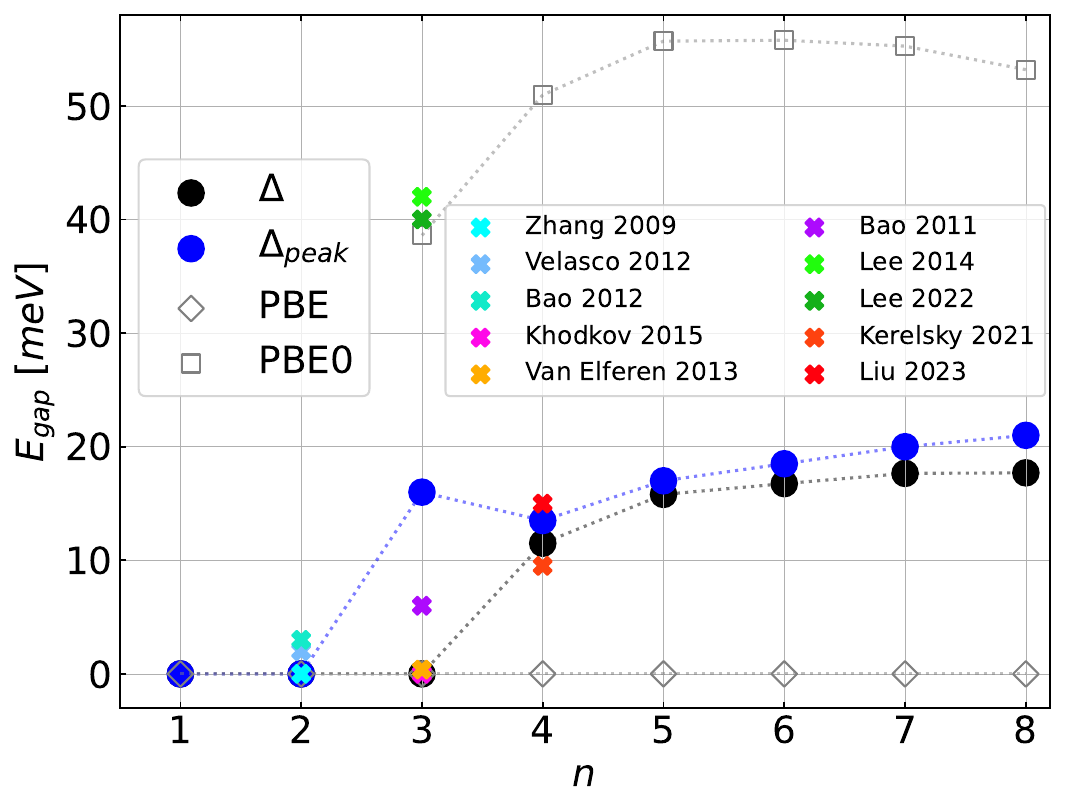}} % !!! refs number need to change !!!
\subfloat[\label{Fig:fig3_d}]{\includegraphics[width=0.325\textwidth]{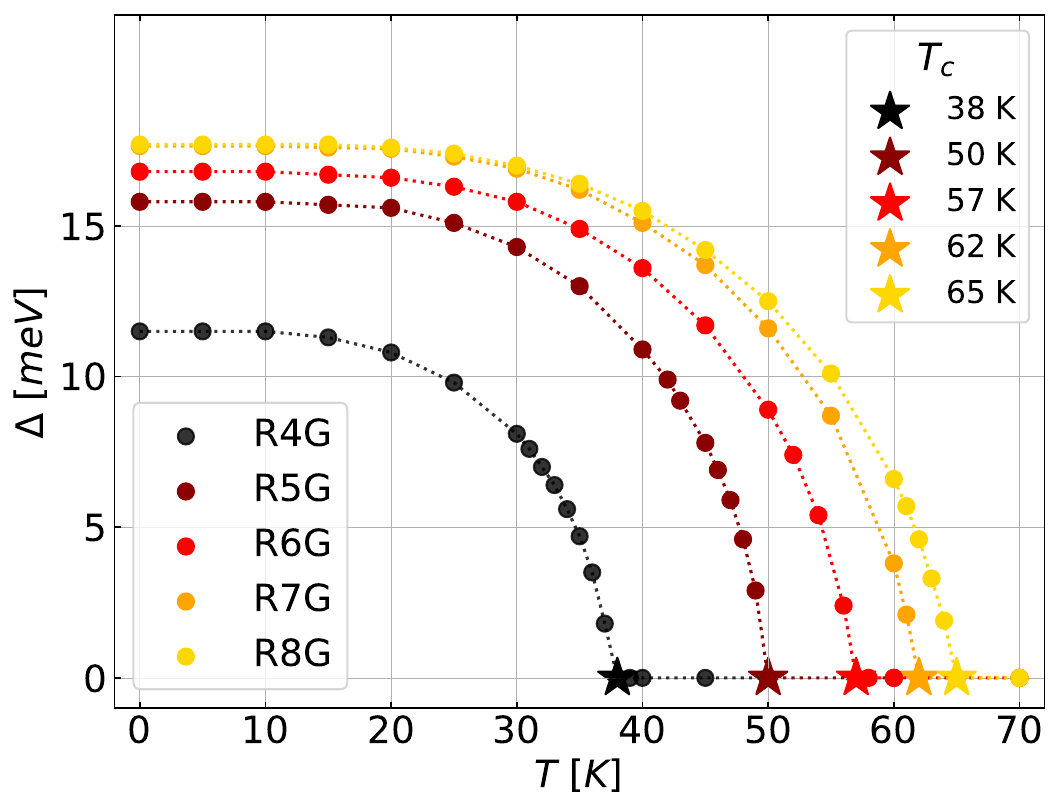}}
\caption{(a) TB+$U$+$V$ band structure and density of states for the ground states of the RnG ($n=3, 4, \dotsb, 8$). The bands show 3D~($k_x,k_y,E$) electronic dispersions projected to 2D~($k_x,E$) plane near the $K$ point. The insets of each panel show the contour plots of the conduction band(left insets) and the valence band(right insets). (b) Bandwidths extracted from the peaks of the density of states. The red straight line is obtained using \Cref{eq:width_fitting}, which is a fitting equation of the valence band flatness for $4 \leq n \leq 8$. (c) Band gaps depending on the number of layers $n$. The gap $\Delta$ represents the difference between the valence band minima and conduction band maxima, while The $\Delta_{peak}$ show the difference in the positions of the two peaks in the density of states. We also list other gaps from PBE(diamond) and PBE0(square)~\cite{pamuk_magnetic_2017,campetella_hybrid-functional_2020}. The cross marks denote experimental band gaps~\cite{khodkov_direct_2015, velasco_transport_2012, bao_evidence_2012, zhang_direct_2009, van_elferen_fine_2013,bao_stacking-dependent_2011,lee_gate-tunable_2022,lee_competition_2014,Kerelsky_moireless_2021, liu_spontaneous_2023}. (d) Temperature dependence of the energy gap $\Delta$. The star marks indicate transition temperatures $T_c$, which represent the lowest temperature points among the data showing zero band gap within an error margin of 1~K.}\label{fig:fig3}
\end{figure*} %<<<< FIG 3 END <<<<%

To validate our model, we compare our calculations for RnG against available experiments.
The \Cref{fig:fig3} presents the electronic dispersions and the density of states~(DOS) near the Dirac $K$ valley of RnG for ($n \leq 8$). 
We do not observe a band gap in R1G, R2G, R3G but see Coulomb-interaction driven finite gaps for $n \geq 4$.
%up to $n=8$ considered. 
%
We first examined the flattening bandwidth $W$ in the vicinity of the Dirac points for quantitative analysis, as shown in \Cref{Fig:fig3_b}. We estimated this bandwidth near the band edges from the full width at half maximum of the DOS peaks,
using the standard deviation $\sigma$ of fitted Gaussian functions to calculate the width as $W = 2\sqrt{2\ln2}\sigma$. The flatness of the conduction band remained nearly constant at around 2~meV for all of the gapped RnG, while the width of the valence band exhibited a linear increase that could be fitted by
\begin{equation} \label{eq:width_fitting}
    W = 1.96 n - 6.52
\end{equation} 
where $n$ is the layer number. Our results can account for the flatness of the experimental results in R4G~\cite{Kerelsky_moireless_2021} and is in keeping with the $\sim$25~meV bandwidth observed in angle-resolved photoemission spectroscopy (ARPES) measurements for R14G~\cite{henck_flat_2018}. 
The Fermi surface geometry also varied notably. While the electron pockets consistently showed a three-fold rotational symmetry along the $K \rightarrow M$ direction for all RnG systems, in contrast the hole pockets showed a similar symmetry for $n \leq 4$, but became annular at R5G and shifted toward the $K \rightarrow \Gamma$ direction for $n \geq 6$. The observed variations in bandwidth and Fermi surface shape provides an explanation for the pronounced electron–hole asymmetry in rhombohedral graphite~\cite{shi_electronic_2020}. The narrow band widths on the order of $\sim$2~meV near the $K$ and $K'$ points in R4G and R5G suggest optimal scenarios for the observation of many-body phenomena.

Next, we examine the size of the band gaps as a function of layer number $n$. Two definitions of band gaps were used in ~\Cref{Fig:fig3_c}. The first one is the proper definition, namely the difference between the conduction band minimum and valence band maximum, denoted as $\Delta$. In our calculations, a finite gap $\Delta$ is found from R4G onwards until it saturates to 18~meV for $n=8$ layers.
The second definition is the density of states (DOS) peak to peak distance $\Delta_{peak}$ between the valence and conduction nearly flat band regions. 
These two values are generally closely similar
%, while some differences can appear due to a progressive increase of the band edge width of the valence band with $n$.
%
with an exception for R3G that shows a particularly strong difference between $\Delta$ and $\Delta_{peak}$. While this $\Delta_{peak}$ could potentially be misinterpreted as a band gap in  spectroscopic measurements, in the case of R3G the gap $\Delta$ remains zero while a finite $\Delta_{peak}$ develops due to the trigonal warping introduced by remote hopping terms that pushes the flattened band regions away from the charge neutral point.   

Experiments in the literature have reported a wide range of results with band gaps spanning from 0 to 42~meV \cite{khodkov_direct_2015, van_elferen_fine_2013, bao_stacking-dependent_2011, zhou_half-_2021, lee_competition_2014, lee_gate-tunable_2022} sensitively depending on experimental conditions such as sample preparation method, choice of substrates or  external electric or magnetic fields. Specifically, the gaps tend to be zero in general when on a substrate even for high quality R3G on misaligned hBN~\cite{zhou_half-_2021}. For R3G on SiO$_2$ a magnetic field induced gap of 0.38~meV was found~\cite{van_elferen_fine_2013}, while in suspended samples the gaps were reported to vary widely, between 0~\cite{khodkov_direct_2015} to 6~meV~\cite{bao_stacking-dependent_2011} or up to 42~meV~\cite{lee_competition_2014}. %Different experimental results were reported in suspended R3G, of either 0~meV~\cite{khodkov_direct_2015} or as large as 42~meV~\cite{lee_competition_2014, lee_gate-tunable_2022} in electronic transport measurements.
%
%
% Except for the case of R3G, the peak to peak distance $\Delta_{peak}$ is closely similar to the gap $\Delta$, but they show some differences primarily due to a progressive increase of the band edge width of the valence band with $n$.
% ------
%
%Our present extended Hubbard calculations predicts a gapless state for R2G~\cite{zhang_direct_2009, cheng_anomalous_2015} unlike earlier predictions of gapped phases in the past literature~\cite{lemonik_spontaneous_2010, vafek_many-body_2010, nandkishore_quantum_2010, zhang_spontaneous_2010, jung_lattice_2011}. 
%
The Coulomb interaction driven band gap of the layer antiferromagnetic phase we found in R4G is of $\Delta \sim 12$~meV, comparable in magnitude with the experimental gap estimates of $\sim$10~meV~\cite{Kerelsky_moireless_2021} and $\sim 15$~meV~\cite{liu_spontaneous_2023}. Previous DFT-based calculations either predicts a zero gap within GGA-PBE~\cite{pamuk_magnetic_2017}, or reaches values of $\sim 50$~meV when calculated through the hybrid PBE0 functional~\cite{campetella_hybrid-functional_2020}, comparable to Hartree-Fock estimates on a tight-binding basis~\cite{jung_gapped_2013}. 
The temperature dependence of the gaps in \Cref{Fig:fig3_d} is introduced through the Fermi-Dirac distribution in the density matrix integration process. The RnG~($n \geq 4$) systems that are gapped transition to gapless metallic phases above their respective critical temperatures $T_c$ of
%. that we define them to be the lowest value resulting in a zero gap. 
38~K for R4G, 50~K for R5G, steadily increasing with layer number $n$ and saturating around 65~K for R8G. 
The value of $T_c = 50$~K for R5G is in keeping with the experimental result in Ref.~\cite{han_correlated_2024} where the correlated insulator phase shows a semi-metallic behavior at charge neutrality.

\begin{figure}[tb] %>>>> FIG 4 >>>>% 
\includegraphics[width=0.95\columnwidth]{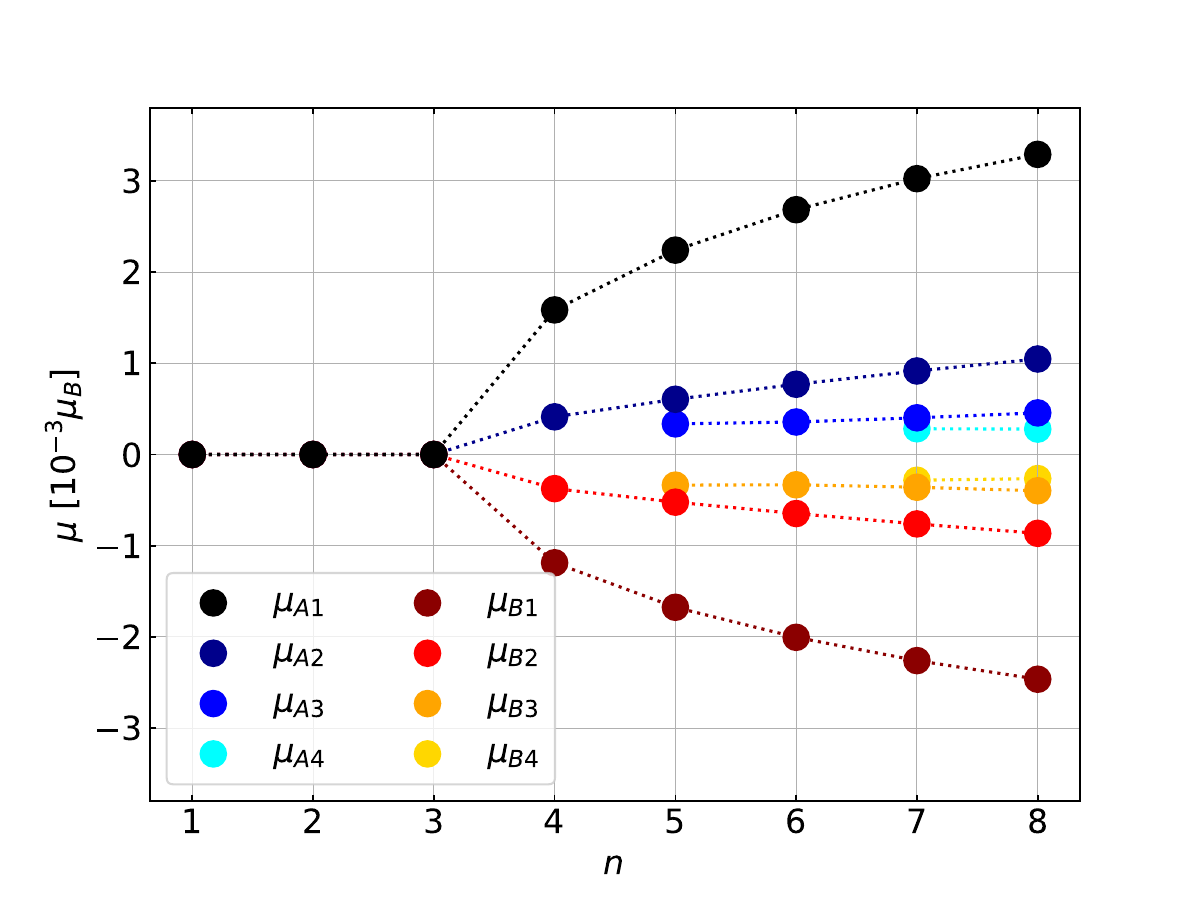}
\caption{Calculated spin magnetic moments of the sublattices in $\mu_B$ as a function of layer number $n$. There is a symmetric relation of the $i^{th}$ layer sublattice with its counterpart from the opposite surface such that $\mu_{Ai} = -\mu_{B{n+1-i}}$.
The sublattice label notation follows \Cref{Fig:fig2}(a).}\label{Fig:fig4}
\end{figure} %<<<< FIG 4 END <<<<%

Stable gapped layer antiferromagnetic (LAF) solutions can be achieved in our calculations calculations for $n \geq 4$, in agreement with earlier work~\cite{myhro_large_2018, Kerelsky_moireless_2021, pamuk_magnetic_2017}. 
We show in \Cref{Fig:fig4} the spin magnetic moments as a function of layer number, where the unlisted sublattices are related by the relation $\mu_{Ai} = -\mu_{B(n+1-i)}$. The magnetic moments are largest at the surface layers and decrease rapidly away from the surface. For a given layer index $i$, 
the $A_i$ and $B_i$ sites exhibit magnetic moments with different signs and magnitudes. Consequently, the net spin magnetic moment of the $i^{th}$ layer is positive for ($i < n/2$) and negative for ($i > n/2$) in the LAF phase. 
%
%This indicates that the origin of the gapped ground states is the layer antiferromagnetic (LAF) phase.
Further discussions on the role of longer range Coulomb interactions in reshasping the DOS peak positions can be found in \Cref{Appendix_UVrange}, where the increase of $\Delta_{peak}$ upon inclusion of longer-ranged  Coulomb interactions can give a misleading appearance of band gap increase in the case of R3G.

\section{Conclusion}\label{conclusion}

In this paper, we introduced the TB+$U$+$V$ model, a self-consistent tight-binding calculation framework to efficiently obtain DFT+$U$+$V$ results using $U$ and $V$ previously obtained from {\em ab initio} calculations. 
Our self-consistent TB+$U$+$V$ calculation for the broken-symmetry solutions uses as the reference non-interacting tight-binding band Hamiltonian the corresponding one for DFT+$U$+$V$ with unbroken spin symmetry, where we subtract the $U$+$V$ contribution.   
%
% --- and the bands are renormalized accordingly. 
%
The increased efficiency of TB+$U$+$V$ has allowed the calculation of the ground states of RnG structures drastically reducing the computation time that would be required in a direct DFT+$U$+$V$ calculation. 
We have applied our model to charge neutral RnG ($n=1, 2, \dotsb, 8$) systems to obtain the extended Hubbard interaction-driven broken symmetry electronic structures. 
We find semimetallic phases for $n \leq 3$ and LAF gapped phases for $n \geq 4$, in excellent agreement with experiments of RnG on hBN susbstrates for the band gap magnitudes and the transition temperatures to the metallic phase.
The results in this work are closer to experiments of RnG systems on substrates than previous calculations in the literature in particular for the reported band gap magnitudes~\cite{jung_tight-binding_2013,schuler_optimal_2013,tancogne-dejean_parameter-free_2020}. 
It is expected that similar TB+$U$+$V$ approaches can be applied for other 2D materials in situations where more efficient computation is required, for instance in typical moir\'e materials with large supercells.

\begin{acknowledgments}
D.L. was supported by the National Research Foundation of Korea (NRF) through grant RS-2024-00413481. Y.-W.S. was supported by KIAS individual Grant (No. CG031509). W.Y. was supported by KIAS individual Grant (No. 6P090103). J.J. was supported by the Basic Study and Interdisciplinary R\&D Foundation Fund of the University of Seoul (2025).
The authors acknowledge the Urban Big Data and AI Institute of the University of Seoul supercomputing resources~(\url{http://ubai.uos.ac.kr}). A part of computations were supported by the CAC of KIAS. 
\end{acknowledgments}

\appendix
\counterwithin{table}{section}
\counterwithin{figure}{section}

\section{Self-consistent tight-binding Extended Hubbard model}\label{Appendix_TBUV}
The Hamiltonian corresponding to the DFT+$U$+$V$ can be cast into the tight-binding Hamiltonian $\hat{H}^0$ in \Cref{Eq:eq_H0}. This Hamiltonian serves as a reference point for the self-consistent TB+$U$+$V$ calculation.
Here, we provide additional details for the derivation of the onsite and intersite hopping terms $\epsilon^{UV}$ and $t^{UV}$ of the Hamiltonian $\hat{H}^0$ stemming from the 
DFT+$U$+$V$ equations presented in Ref.~\cite{Timrov2018PRB}. 
%In our calculations we have used the self-consistent $U$ and $V$ parameters following Lee-Son Ref.~\cite{lee_first-principles_2020}, that is similar to the Tancogne-Dejean-Rubio~\cite{tancogne-dejean_parameter-free_2020}, that provides a recipe to calculate the $V$ parameter values following the method proposed Agapito-Curtaolo-Buongiorno-Nardelli in Ref.~\cite{agapito_reformulation_2015} for the calculation of the $U$ values. 
%
%

We use the localized orbital basis  $\phi_i(\mathbf{r})$ at atomic $i$ sites for our tight-binding model. Hence, the density operators 
    $\hat{\rho}(\mathbf{r}) = \hat{\psi}^\dagger(\mathbf{r}) \hat{\psi}(\mathbf{r})$
such that the field operators
    $ \hat{\psi}^\dagger(\mathbf{r}) = \sum_i \phi_i^*(\mathbf{r}) \hat{c}_i^\dagger $
    and     $\hat{\psi}(\mathbf{r}) = \sum_j \phi_j(\mathbf{r}) \hat{c}_j $
are defined in terms of atomic orbital creation and annihilation operators. 

The Kohn-Sham (KS) DFT+$U$+$V$ Hamiltonian
\begin{equation}
    \hat{H}^{\text{DFT}+U+V} = \sum_{i,j} h_{ij}^{\text{KS}} \hat{c}_i^\dagger \hat{c}_j
\end{equation}
consists of matrix elements
\begin{align*}
    &h_{ij}^{\text{KS}}  \\ &= \int d^3r \, \phi_i^*(\mathbf{r}) \left[ -\frac{\hbar^2}{2m} \nabla^2 + V^{\text{DFT}}[\rho](\mathbf{r})+ V^{UV}[\rho](\mathbf{r}) \right] \phi_j(\mathbf{r})
\end{align*}
where the kinetic term gives rise to the DFT interatomic hopping terms, and the KS potential is
\begin{align}
    V^{\text{DFT}}[\rho]({\bf r}) 
    &=  
   V_{\rm ext}(\mathbf{r}) + V^{\text{H}}[\rho] (\mathbf{r}) + v_{\text{xc}}[\rho](\mathbf{r}).
\end{align}
Apart from the standard KS terms a widely used way to deal with the double-counting corrected form of the $U$+$V$  potential\cite{leiria_campo_jr_extended_2010} is the so-called ‘fully localized limit’ (FLL). In the restriction for the single $p_z$ orbitals at the carbon atoms

% Full form in Leiria-Coccocione, we remove the sum in m and m' in our case
%\begin{align}
%    \hat{V}^{UV}
%    &=  \sum_{i} \frac{U_i}{2} \sum_{m m^{\prime}} \left( \delta_{m m^{\prime}} - 2 \rho^{i i \sigma}_{m^{\prime} m}\right)  \hat{c}^{\dagger}_{i \sigma} \hat{c}_{i \sigma} \\
%    &- \sum_{i \neq j} V_{ij} \sum_{m m^{\prime}} \rho^{j i \sigma}_{m^{\prime} m} \hat{c}^{\dagger}_{j \sigma} \hat{c}_{i \sigma}.
%\end{align}
\begin{align*}
    \hat{V}^{UV}
    =& \hat{V}^{\text{Hub}} - \hat{V}^{\text{dc}}\\
    =&  \sum_{i, \sigma} \frac{U_i}{2}  \left( 1 - 2 \rho^{\sigma }_{i i}  \right)  \hat{c}^{\dagger}_{i \sigma} \hat{c}_{i \sigma}\\
    &- \sum_{\langle i, j \rangle, \sigma} V_{ij}  \rho^{\sigma }_{i j } \hat{c}^{\dagger}_{i \sigma} \hat{c}_{j \sigma}, \nonumber
\end{align*}
where $\langle i, j \rangle$ represents sums over all possible unequal sites within a given truncation range.
The inter-site interactions $V_{ij}$ are assumed to be the same as the
Coulomb interaction averaged over all the states, just one orbital per atomic site in our case, such that 
% Form in Leiria and Coccocione. In their notation 
%
% i,j comprehensive state index, i.e. i are atomic orbitals of atom I
% I,J site index.  
% m is the magnetic quantum number
% In our case we do not have sums in m and neither for I,J  
%
%$\langle \phi^{I}_{i} \phi^{J}_{j} \mid V_{ee} \mid \phi^{K}_{k} \phi^{L}_{l} \rangle \rightarrow 
%\delta_{IK} \delta_{JL} \delta_{ik} \delta_{jl} V^{IL} =
%\frac{\delta_{IK} \delta_{JL} \delta_{ik} \delta_{jl} }{(2l_I + 1)(2l_{J} + 1)}
%\sum_{i^{\prime} j^{\prime}}
%\langle \phi^{I}_{i^{\prime}} \phi^{J}_{j^{\prime}} \mid V_{ee} \mid \phi^{K}_{k^{\prime}} \phi^{L}_{l^{\prime}} \rangle
$
V_{ij} = \langle \phi_{i} \phi_{j}\mid V_{ee} \mid \phi_{i} \phi_{j} \rangle.
$ The same consideration applies when incorporating the Hubbard correction into hopping parameters obtained from DFT~\cite{hourahine_self-interaction_2007}.

In contrast, in this work, we employ hopping parameters obtained from DFT+U+V calculations rather than those obtained from the conventional DFT. By utilizing the parameters from which the double-counting interaction contained in $V^{\text{DFT}}$ has been eliminated, we ensure that only the Hubbard-model-like potential term $V^{\text{Hub}}$ corresponding to the variations in the density matrix need to be considered in subsequent tight-binding calculations. The DFT+U+V potential can be reformulated by decomposing it into a density-independent non-interacting part~$\hat{V}^{\text{NI}}$, an onsite interaction~$\hat{V}^{\text{on}}$, and an intersite interaction~$\hat{V}^{\text{inter}}$.
\begin{align*}
    \hat{V}^{\text{DFT}+U+V}[\rho]
    =& \hat{V}^{\text{DFT}}[\rho] +\hat{V}^{\text{Hub}}[\rho] - \hat{V}^{\text{dc}}[\rho]\\
    =& \hat{V}^{\text{NI}} +\hat{V}^{\text{on}}[\rho] + \hat{V}^{\text{inter}}[\rho]
\end{align*}
In DFT+U+V, the onsite component of $V^{\text{DFT}}$ is effectively removed by the onsite $V^{\text{dc}}$, such that the remaining onsite term $V^{\text{on}}$ corresponds to the Hubbard U term of the Hubbard potential $V^{\text{Hub}}$. Similarly, since the intersite $V^{\text{dc}}$ corresponds to the Hartree part of $V^{\text{Hub}}$, the total intersite interaction can be expressed as the interaction part of $V^{\text{DFT}}$, equally $V^\text{H} + v_{xc}$, supplemented by the Fock (exchange) part of $V^{\text{Hub}}$:
\begin{align*}
    \hat{V}^{\text{on}}[\rho] &= \sum_i U_i \rho^{\tilde \sigma}_i \hat{c}^{\dagger}_{i \sigma} \hat{c}_{i \sigma}\\
    \hat{V}^{\text{inter}}[\rho] &= (\sum_{ij} V^H_{ij}\rho_j +v_{xc,i\sigma}) \hat{c}^\dagger_{i\sigma} \hat{c}_{i\sigma} -\sum_{ij\sigma}V_{ij} \rho^\sigma_{ij}\hat{c}^\dagger_{i\sigma} \hat{c}_{j\sigma} 
\end{align*}
Based on this, the tight-binding parameters from the DFT+U+V can be derived as follows:
\begin{align}
\epsilon^{UV}_{i\sigma} :=& h^{\text{KS}}_{ii\sigma} \nonumber \\
=& \epsilon^{\text{NI}}_{i\sigma} + U_i \rho_{i\tilde\sigma} + \sum_j V^{\rm H}_{ij} \rho_j + v_{xc,i\sigma},\\
t^{UV}_{ij\sigma} :=& h^{\text{KS}}_{ij\sigma} \nonumber\\
=& t^{\text{NI}}_{ij\sigma} - V_{ij} \rho_{ij\sigma},
\end{align}
where $\epsilon^{\text{NI}}$ and $t^{\text{NI}}$ represent the non-interacting parts of the onsite and hopping parameters, respectively. These are obtained from the kinetic term and the non-interacting potential~$V^{\rm NI}$.

\section{Hopping range truncation and simplified effective tight-binding model}\label{Appendix_FnGn}

\begin{table}[bt!] %>>>> TABLE >>>>%
\begin{threeparttable}
\caption{Hopping parameters of the simple model for RnG in $eV$. This model is defined by hopping parameters $t^{\Delta l}_{\alpha\beta}$ for the layer difference $\Delta l$ and sublattices $\alpha,\beta$, along with diagonal site potentials $u_{\alpha}$~[For an example of application to R4G, see \Cref{Fig:figB1_b}]. }\label{tab:simplemodel}
\begin{ruledtabular}
\def\arraystretch{1.3}%  1 is the default, change whatever you need
\begin{tabular}{cccc}
$u_{A_1}/u_{B_n}$ & $u_{B_1}/u_{A_n}$ & $u_{other}$ & $t^0_{AB}$ \\
\hline
0.018 & 0.067 & 0.0 & $-$3.415\\
\hline \hline
$t^1_{AA}/t^1_{BB}$ & $t^1_{AB}$ & $t^1_{BA}$ & \\
\hline
0.161 & 0.279 & 0.343 & \\
\hline \hline
$t^2_{AA}/t^2_{BB}$ & $t^2_{AB}$ & $t^2_{BA}$ & \\
\hline
0.009 & $-$0.009 & 0.019 & \\
\end{tabular}
\end{ruledtabular}
\end{threeparttable}
\end{table} %>>>> TABLE-END >>>>%

\begin{figure*}[bth!] %>>>> FIG A1 >>>>% 
\subfloat[\label{Fig:figB1_a}]{\begin{tabular}{c}
\includegraphics[width=0.300\textwidth]{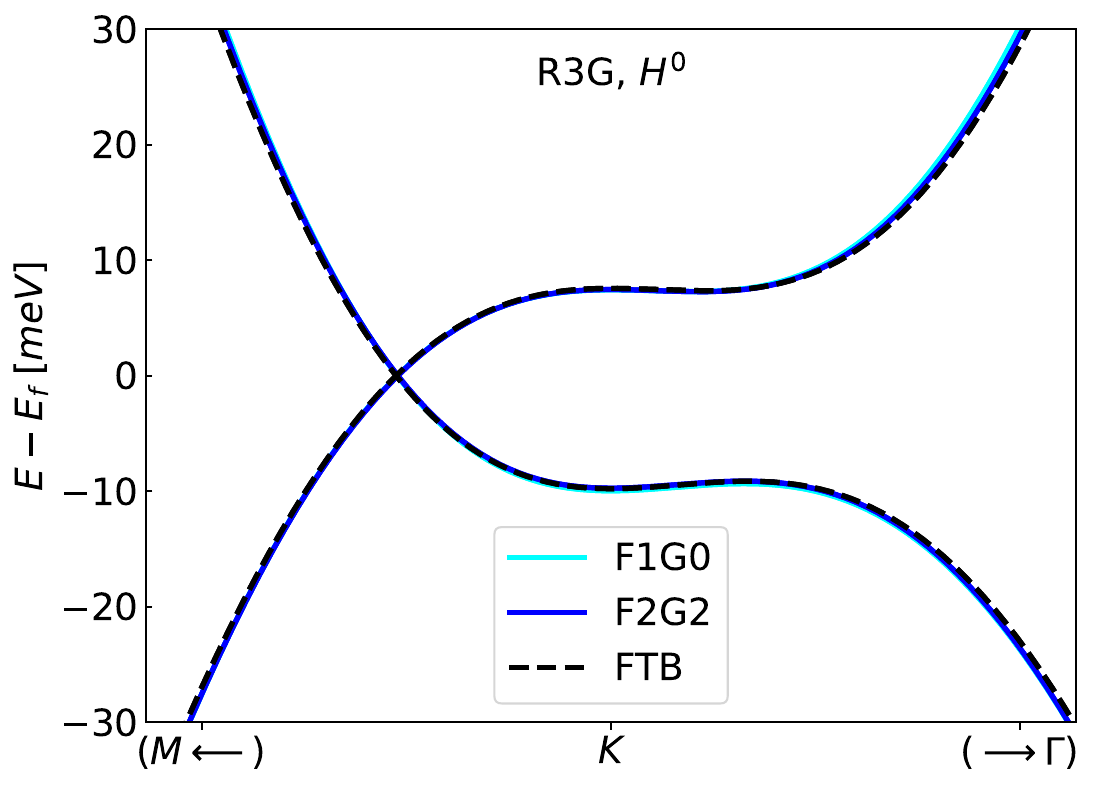}
\includegraphics[width=0.300\textwidth]{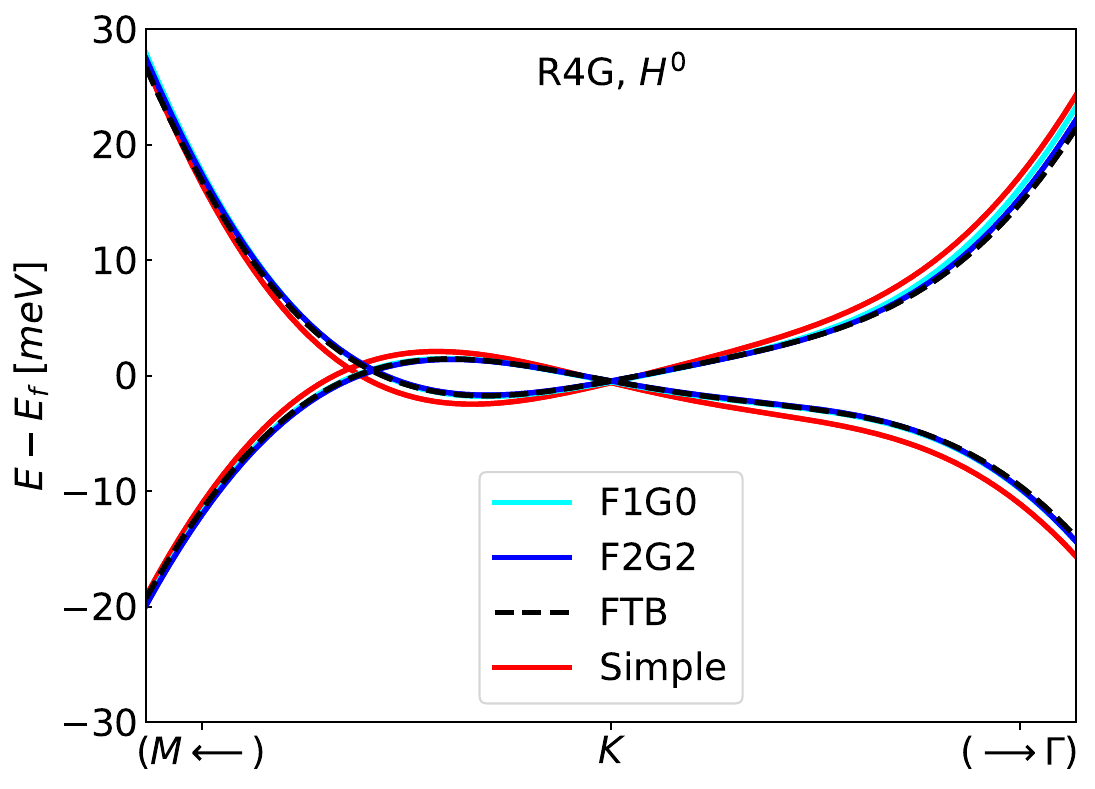}\\
\includegraphics[width=0.300\textwidth]{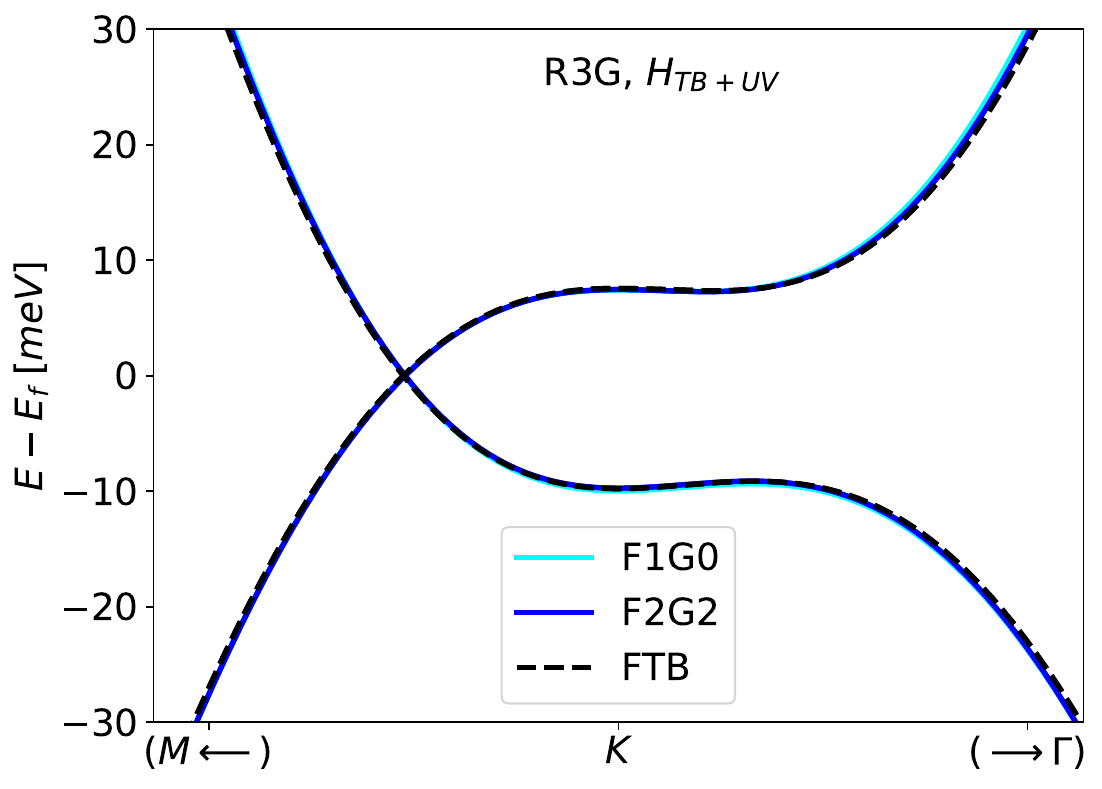}
\includegraphics[width=0.300\textwidth]{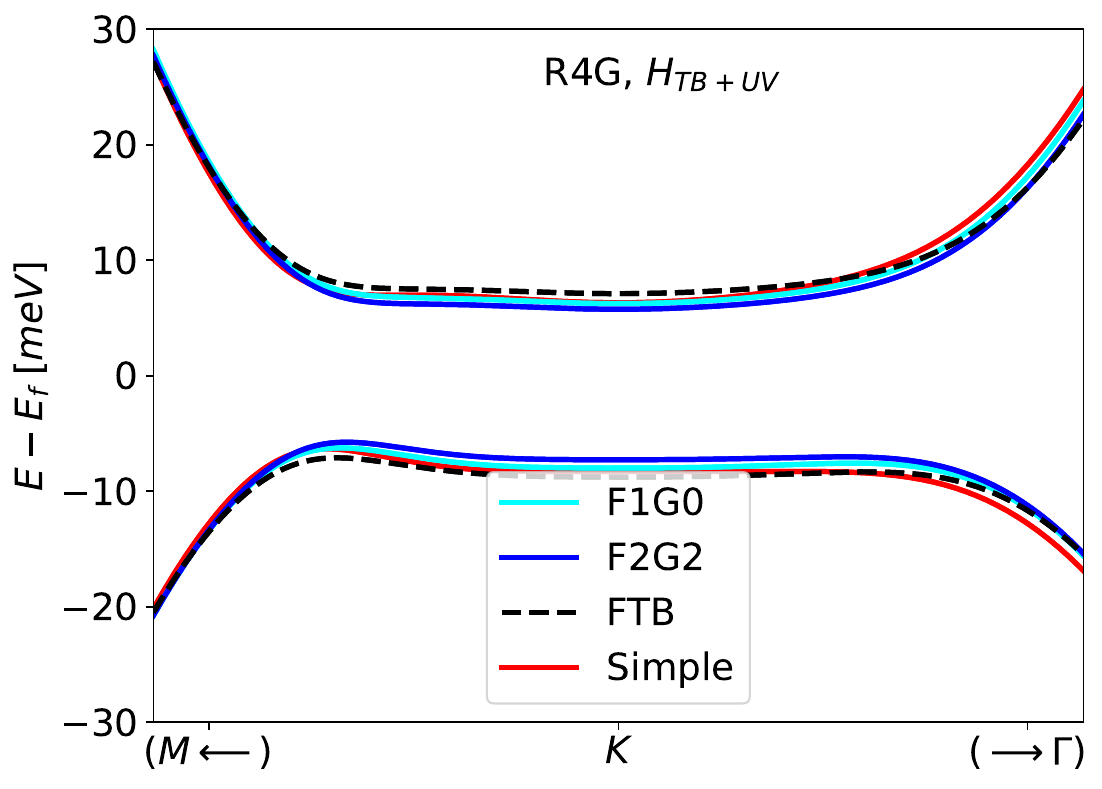}
\end{tabular}}
\subfloat[\label{Fig:figB1_b}]{\includegraphics[width=0.33\textwidth]{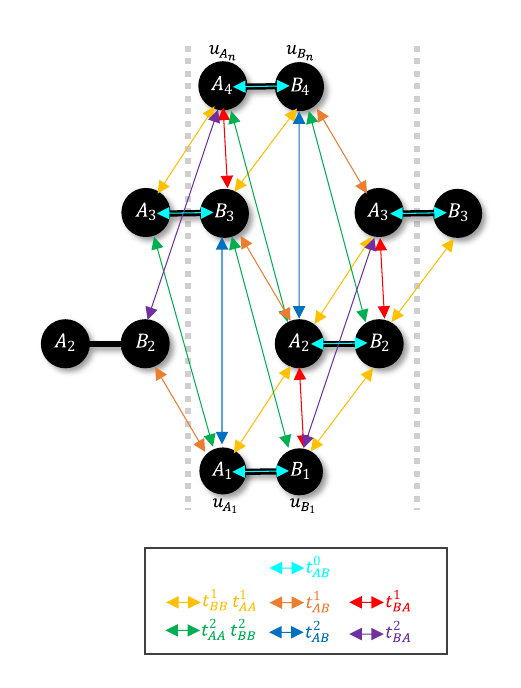}}
\caption{ (a) Comparison of energy dispersions near the $K$ point for $\hat{H}^0$ and self-consistent $\hat{H}^{{\rm TB}+U+V}$ in R3G and R4G using different models. Models shown include full tight-binding(FTB) model, the truncated models with F1G0 and F2G2, and the simple model in \Cref{tab:simplemodel}. (b) Schematic representation of hopping processes when applying our simple model to R4G.}\label{fig:figB1}
\end{figure*} %<<<< FIG A1 END <<<<%

The $F_nG_n$ truncation~\cite{jung_accurate_2014,jung_tight-binding_2013} determines the hopping model to preserve the zeroth and first-order $\vec{k} \cdot \vec{p}$ expansion coefficients in the vicinity of the Dirac cone. For a truncation order $n$, the effective hopping parameters $t$ of orders less than $n$ are retained identical to those of the full tight-binding(FTB) model $t^{\text{FTB}}$, while the $n$-th order truncated hopping coefficient is given by:
\begin{equation}
t_{\alpha\beta,n} = \frac{\sum_{m=n}^\infty c^{f/g}_m t^{\text{FTB}}_{\alpha\beta,m}}{c^{f/g}_n}
\end{equation}
where $\alpha$ and $\beta$ denote the sublattice indices, and the coefficients $c^{f/g}$ are associated with the $F/G$ categories and are derived from the structure factors at the Dirac point.
For the $F$ category, which encompasses sublattice pairs with distinct planar positions (e.g., $\alpha\beta = A_1 B_1, A_1A_2$), the coefficients are given by:
\begin{equation}
    (c^f_1, c^f_2, \dots) = (-1, +2, +1, -5, -4, +7,  \dots).
\end{equation}
Conversely, when two sublattices occupy same planar positions, exemplified by $\alpha\beta = A_1A_1, A_1B_3, \dots$, they fall under the $G$ category. For the cases, the coefficients are 
\begin{equation}
    (c^g_0, c^g_1, \dots) = (1, -3, +6, -3, -6, +6, \dots).
\end{equation}
The $t_0$ term in the $G$ category represents the onsite potential for $\alpha = \beta$, while for $\alpha \neq \beta$, it denotes the perpendicular hopping.

We validated the $F_nG_n$ truncation for RnG systems. We examined the energy dispersion near the Dirac point for R3G and R4G using the FTB model, as well as the $F_2G_2$ and $F_1G_0$ models for both the $\hat{H}^0$ and the $\hat{H}^{{\rm TB}+U+V}$. As illustrated in \Cref{Fig:figB1_a}, we confirmed that not only $F_2G_2$ but also $F_1G_0$ terms listed in \Cref{tab:simplemodel} can significantly simplify the model without substantial loss of accuracy, both for $\hat{H}^0$ and the self-consistent $\hat{H}^{{\rm TB}+U+V}$. Furthermore, we observed that applying the simplified parameters of the $F_1G_0$ model for R3G to R4G resulted in no significant differences in band structure or ground state near the Dirac point and therefore can be used for all RnG models. 
The full tight-binding parameters in HR-format of {\textsc{Wannier90}}~\cite{pizzi_wannier90_2020} for all RnG systems, along with a Python script for performing $F_nG_n$ truncation, are available in our GitHub repository~\cite{hopping_github}.

\begin{figure*}[bt!] %>>>> FIG B1 >>>>% 
\includegraphics[width=0.30\textwidth]{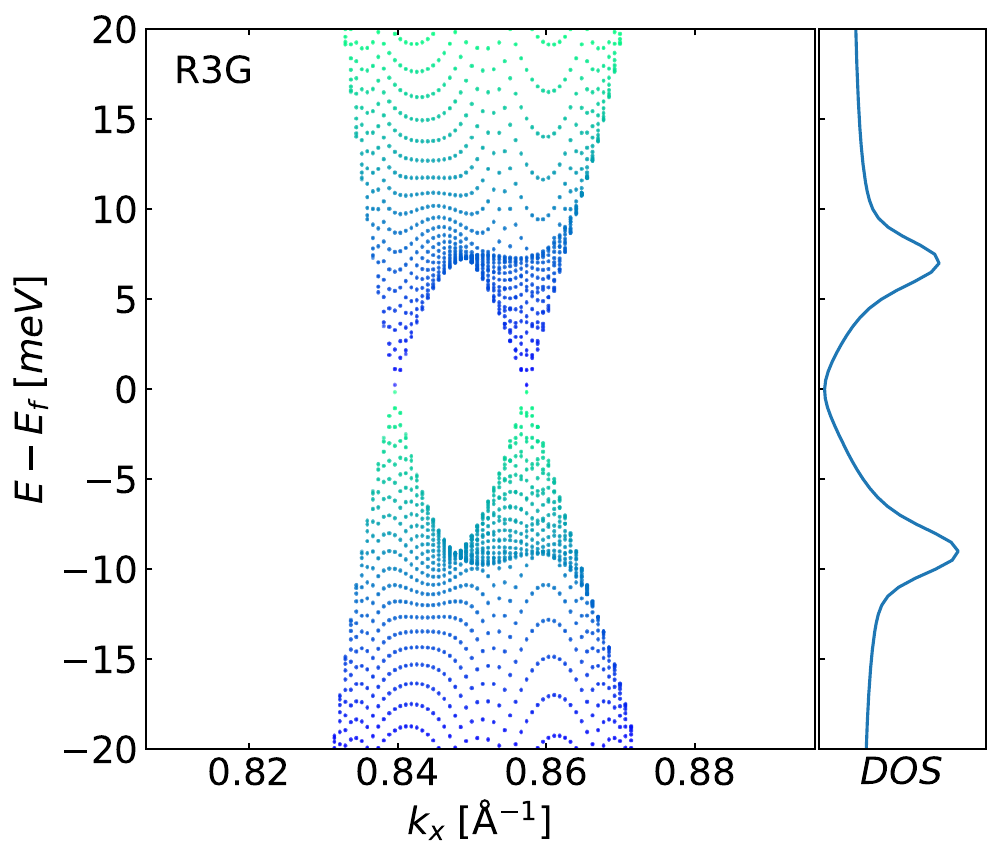}
\includegraphics[width=0.30\textwidth]{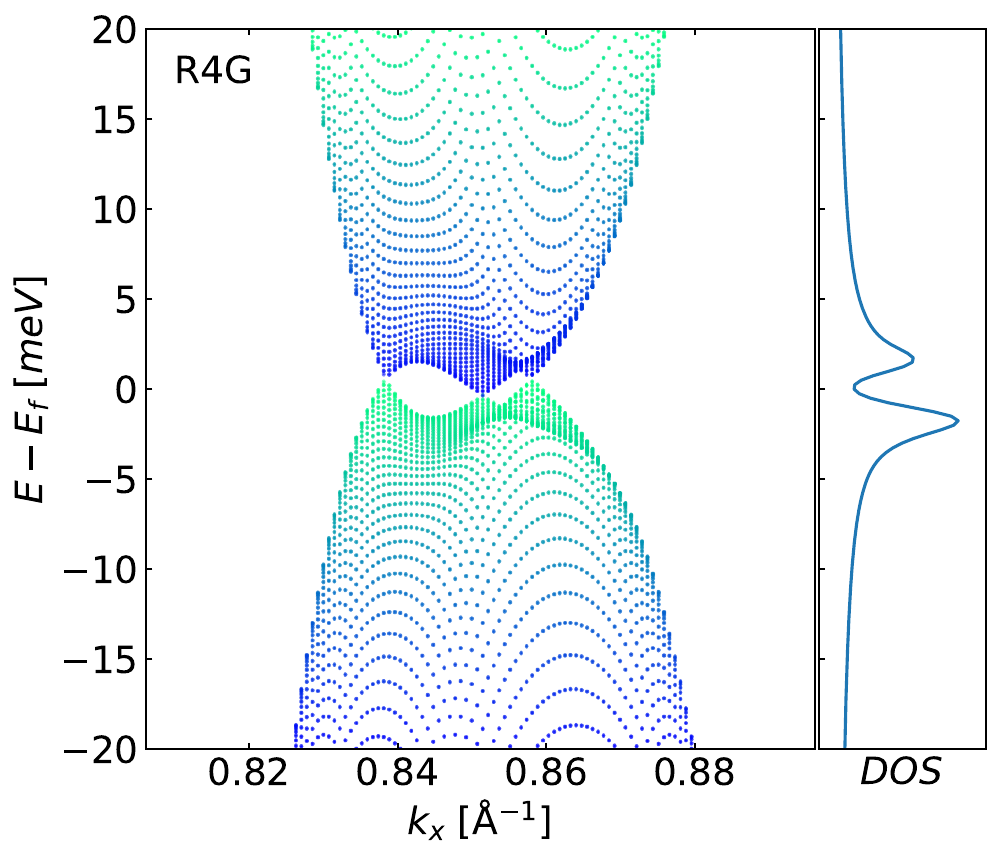}
\includegraphics[width=0.30\textwidth]{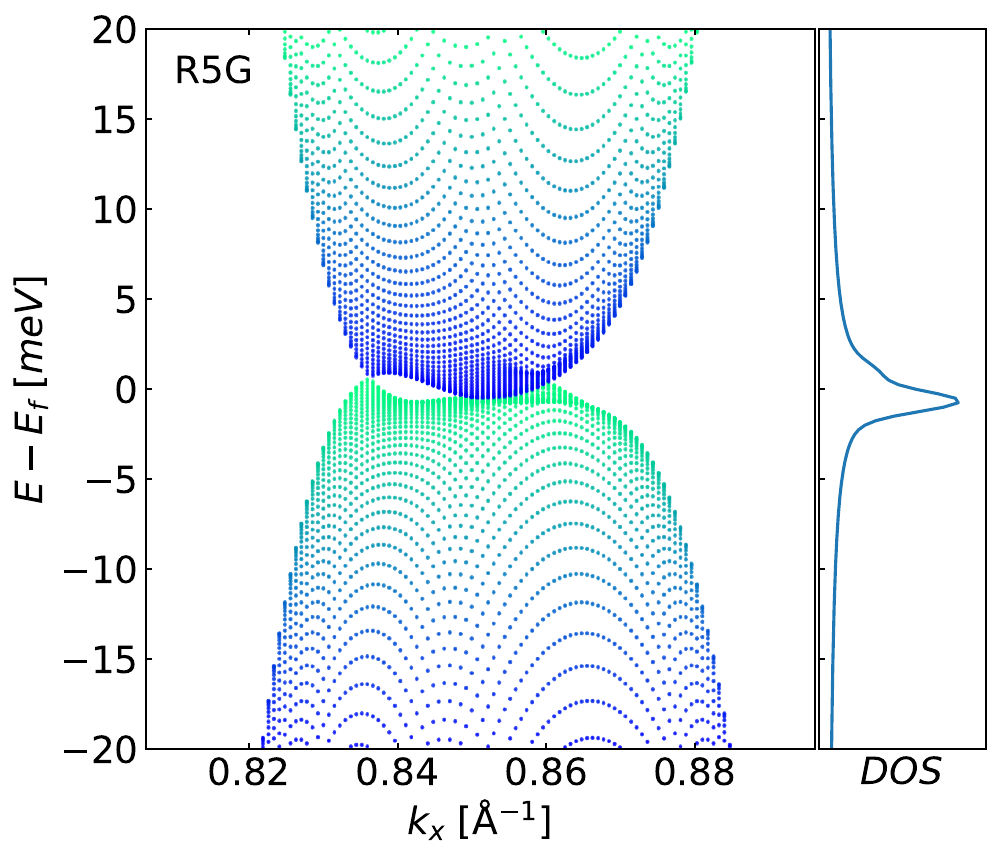}
\\
\includegraphics[width=0.30\textwidth]{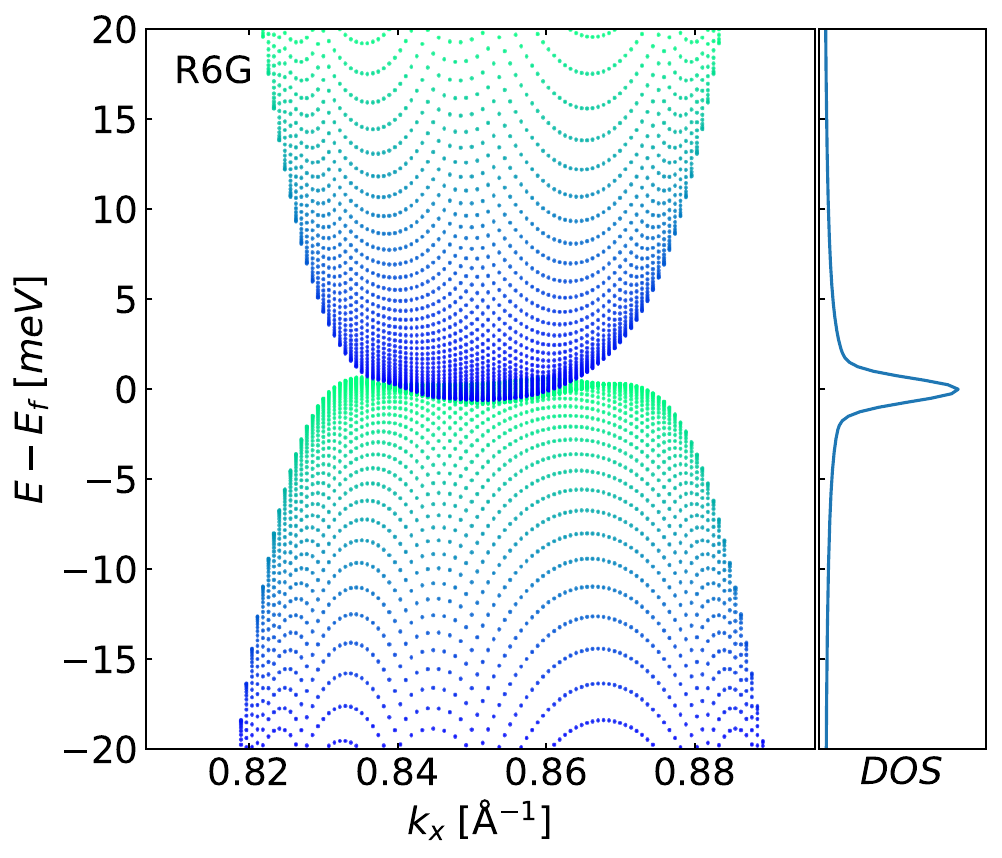}
\includegraphics[width=0.30\textwidth]{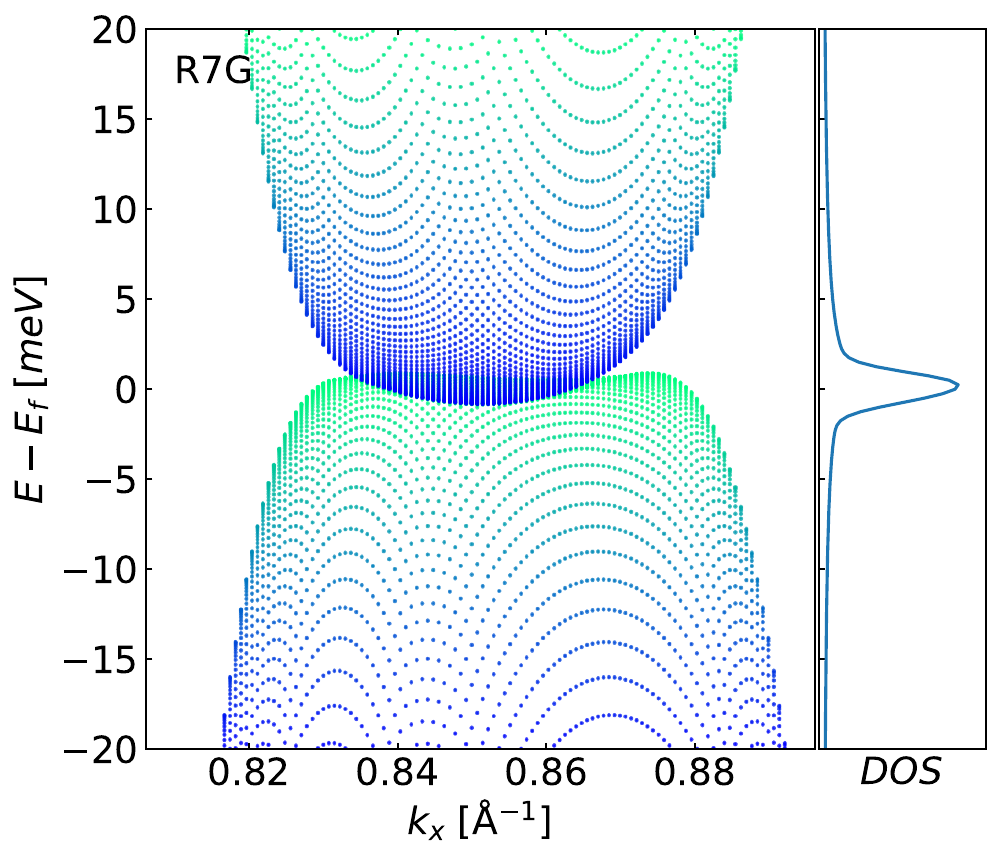}
\includegraphics[width=0.30\textwidth]{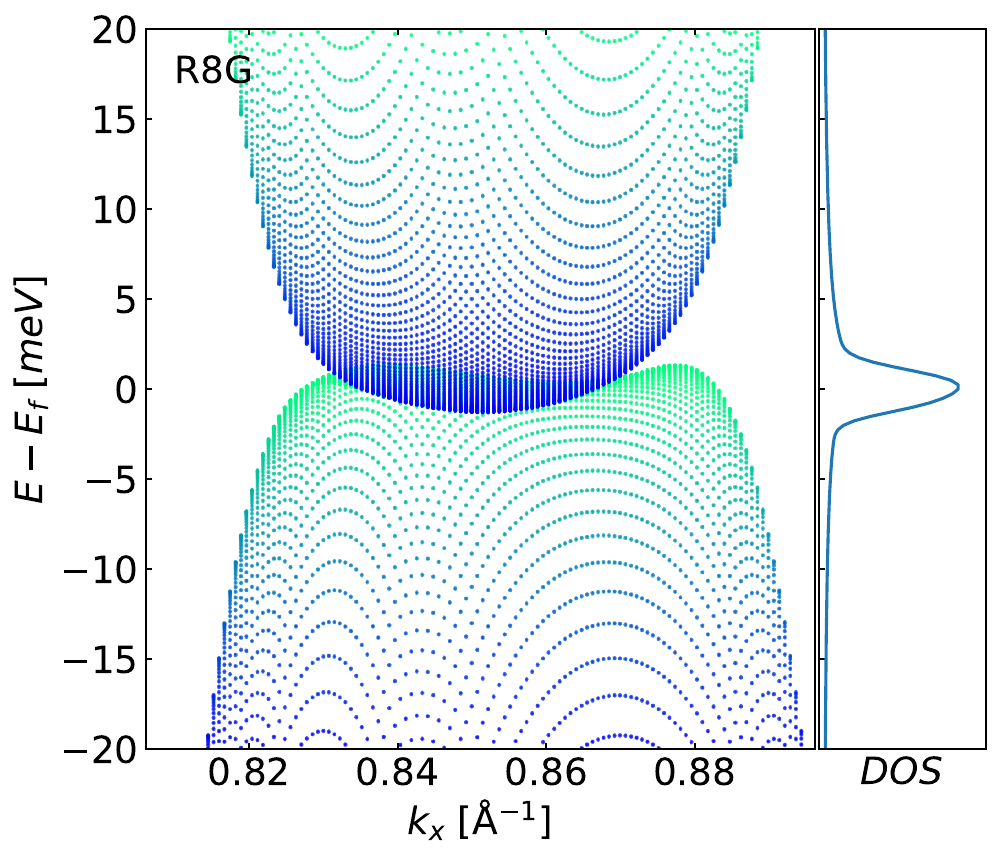}
\caption{$\hat{H}^0$ band structure near the $K$ point and density of states of the $F_2G_2$ models for the RnG ($n=3, 4, \dotsb, 8$).}\label{fig:figB2}
\end{figure*} %<<<< FIG B1 END <<<<%

The non-corrected Hamiltonian $\hat{H}^0$ in \Cref{Eq:eq_H0} yields the same band structure as in the DFT+$U$+$V$ step, representing the ground state under spin non-polarized conditions. Consequently, for R1G, R2G, and R3G, which exhibit non-magnetic ground states in $\hat{H}^{{\rm TB}+U+V}$, the band structures of $\hat{H}^0$ and $\hat{H}^{{\rm TB}+U+V}$ are identical. \Cref{fig:figB2} show the electronic band and DOS of the $F_2G_2$ model of the $\hat{H}^0$ for RnG.

\section{Role of Coulomb correction cutoff in the extended Hubbard model}\label{Appendix_UVrange}

\begin{figure*}[bt!] %>>>> FIG 5 >>>>% 
\subfloat[\label{Fig:figC1_a}]{\includegraphics[width=0.85\columnwidth]{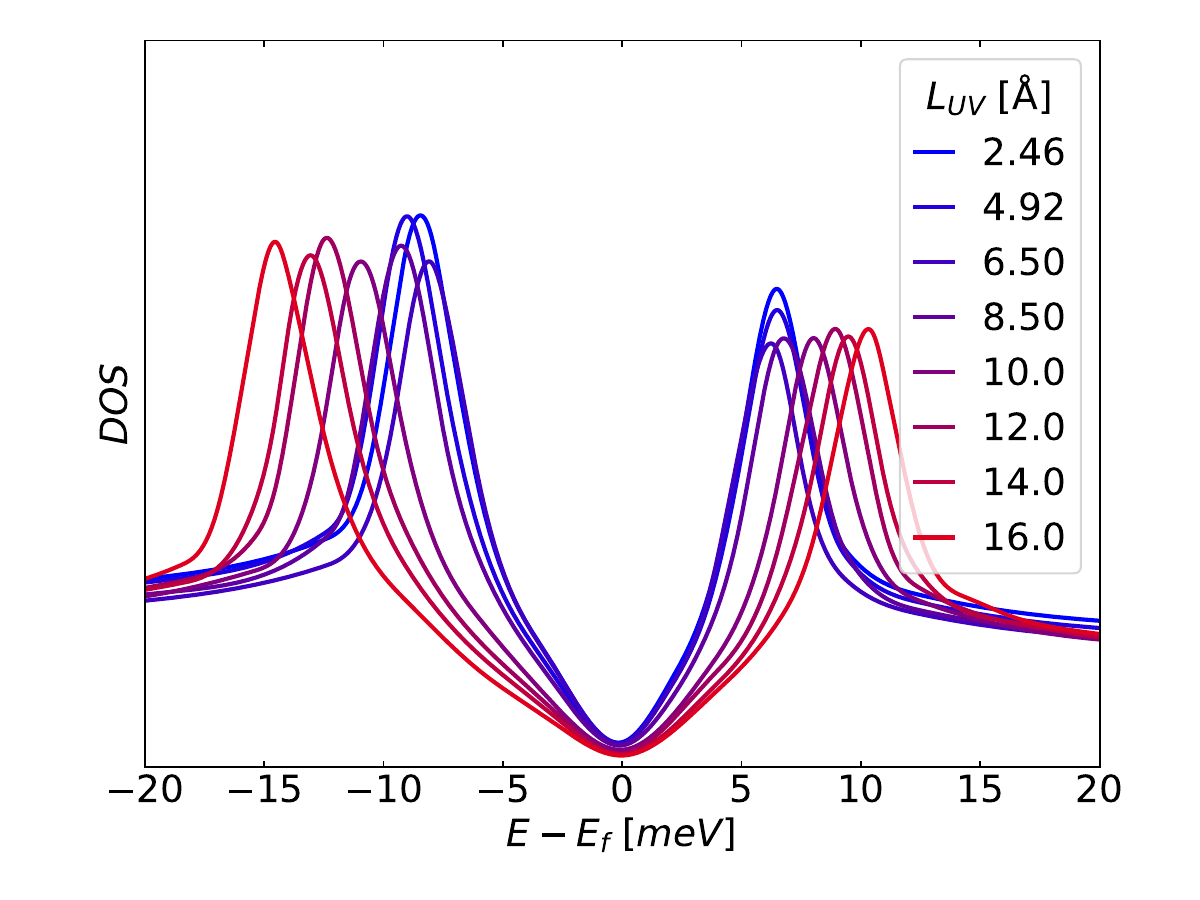}}
\subfloat[\label{Fig:figC1_b}]{\includegraphics[width=0.85\columnwidth]{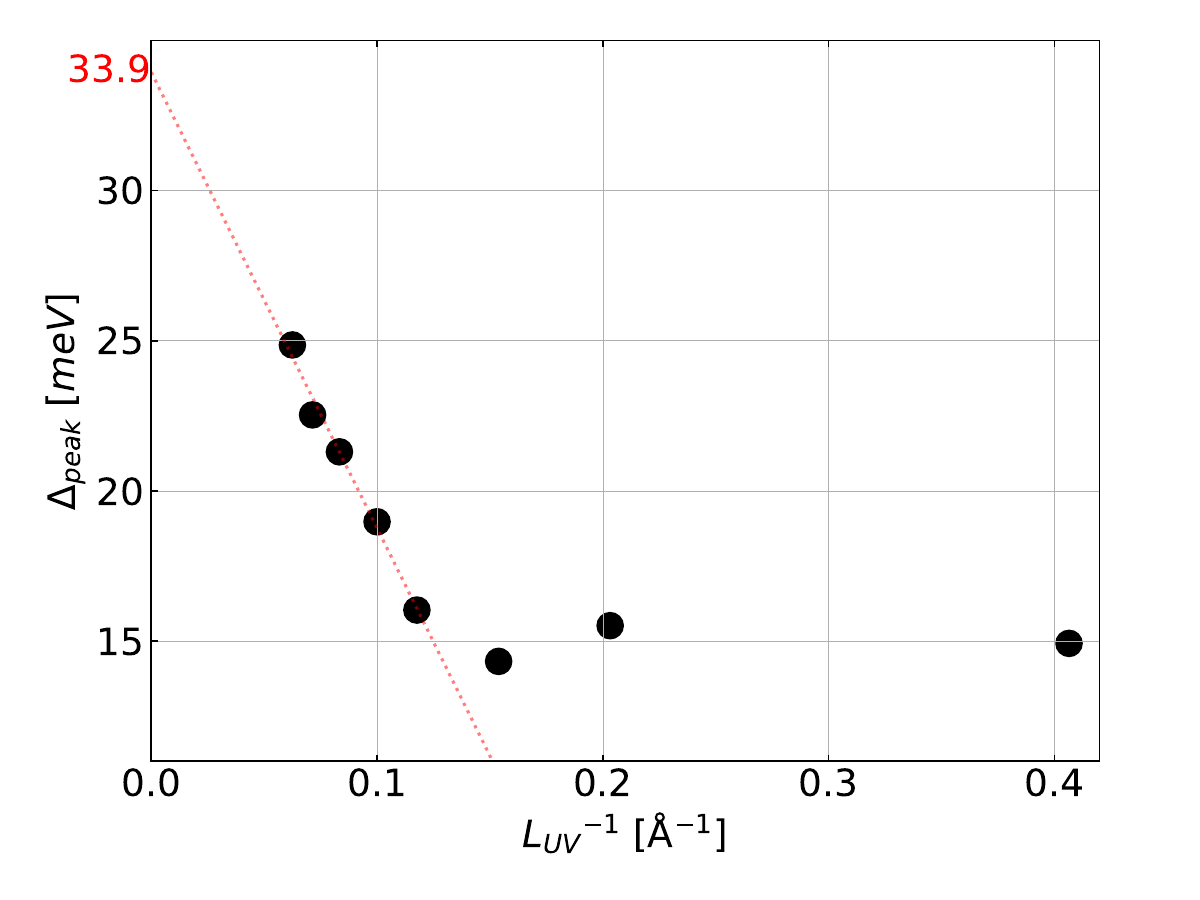}}
\caption{(a) Comparison of the density of states of TB+$U$+$V$ models in rhombohedral trilayer graphene with varying correction ranges $L_{UV}$.  For this calculation we have used all the distant hopping parameters in the band Hamiltonian. 
(b) Interaction range dependence of the peak to peak DOS distance $\Delta_{peak}$. 
In systems with long-range Coulomb interactions, where screening is suppressed, the separation between DOS peaks becomes larger, which can give the misleading appearance of an increasing band gap. The red dashed line fitting the data for $L_{UV} \geq 8.5$ exhibits a $y$-intercept of 33.9 meV. }\label{Fig:figC1}
\end{figure*} %<<<< FIG 5 END <<<<%

In our calculation we have truncated the range of the Coulomb interaction up to the $V_2$ term for most of the reported results. 
Here we discuss the evolution of the band structure shapes in R3G for increasing Coulomb interaction range in our TB+$U$+$V$ calculation.
%that is neglected due to the truncation up to $V_2$ in plane term. 
It is known that the Fermi velocity of single layer graphene depends on the range of the Coulomb interactions~\cite{elias_dirac_2011}, and it is also expected to reshape the band structure of R3G with increasing $L_{UV}$ where the absence of the band gap $\Delta$ and increased trigonal warping leads zero and finite values of the gap and the peak to peak distance of the DOS $\Delta_{peak}$ respectively.
\Cref{Fig:figC1} shows the DOS obtained from our model for increasing extended Hubbard correction cutoff $L_{UV}$. In this calculation we used the full tight-binding hopping parameters for the band Hamiltonian instead of the $F_2G_2$ truncation. 
%Here, we bring to discussion the case of R3G that in our truncated range Coulomb model does not have a band gap, while numerous experiments have reported sizeable gaps of up to a few meVs~\cite{van_elferen_fine_2013, bao_stacking-dependent_2011,lee_competition_2014, lee_gate-tunable_2022}. Specifically, suspended R3G~\cite{lee_competition_2014, lee_gate-tunable_2022} tends to exhibit a larger bandgap compared to R3G on substrates~\cite{van_elferen_fine_2013, bao_stacking-dependent_2011}. 
% ---
%A plausible conciliation of the experimental observation with our calculation can be in the fact that 
%
%
%It is known that 
%Similarly to the divergence of the Fermi velocity observed in 
%monolayer graphene's Fermi velocity diverges for infinite range Coulomb interactions~\cite{elias_dirac_2011}, 
The main effect we find is the reduction of the DOS near the Fermi level and increase of  $\Delta_{peak}$.
We extrapolate the $\Delta_{peak}$ value to the limit of infinite range by using a linear fit with respect to $L_{UV}^{-1}$ for some values of $L_{UV} \geq 8.5~\text{\AA}$.
%, {\bf that is greater than the distance between $A_1$ and $B_3$}. 
This extrapolation from finite-size scaling in \Cref{Fig:figC1_b} yields a $y$-intercept of 33.9~meV. 

\newpage
\bibliography{aps}% Produces the bibliography via BibTeX.

\end{document}